\documentclass{emulateapj}

\shorttitle{High-redshift mid-IR Galaxy Cluster Luminosity Function}
\shortauthors{Wylezalek et al.}

\begin{document}

\title{The Galaxy Cluster Mid-Infrared Luminosity Function at $1.3<z<3.2$}

\author{Dominika Wylezalek\altaffilmark{1}, Jo\"{e}l Vernet\altaffilmark{1}, Carlos De Breuck\altaffilmark{1}, Daniel Stern\altaffilmark{2}, Mark Brodwin\altaffilmark{3}, Audrey Galametz\altaffilmark{4}, Anthony H. Gonzalez\altaffilmark{5}, Matt Jarvis\altaffilmark{6,7}, Nina Hatch\altaffilmark{8},  Nick Seymour\altaffilmark{9}, Spencer A. Stanford\altaffilmark{10, 11}} 

\altaffiltext{1}{European Southern Observatory, Karl-Schwarzschildstr.2, D-85748 Garching bei M\"{u}nchen, Germany}
\altaffiltext{2}{Jet Propulsion Laboratory, California Institute of Technology, 4800 Oak Grove Dr., Pasadena, CA 91109, USA}
\altaffiltext{3}{Department of Physics and Astronomy, University of Missouri, 5110 Rockhill Road, Kansas City, MO 64110, USA}
\altaffiltext{4}{INAF - Osservatorio di Roma, Via Frascati 33, I-00040, Monteporzio, Italy}
\altaffiltext{5}{Department of Astronomy, University of Florida, Gainesville, FL 32611, USA}
\altaffiltext{6}{Astrophysics, Department of Physics, Keble Road, Oxford OX1 3RH, UK}
\altaffiltext{7}{Physics Department, University of the Western Cape, Bellville 7535, South Africa}
\altaffiltext{8}{School of Physics and Astronomy, University of Nottingham, University Park, Nottingham, NG7 2RD, UK}
\altaffiltext{9}{CASS, PO Box 76, Epping, NSW, 1710, Australia}
\altaffiltext{10}{Physics Department, University of California, Davis, CA 95616, USA}
\altaffiltext{11}{Institute of Geophysics and Planetary Physics, Lawrence Livermore National Laboratory, Livermore, CA 94550, USA}

\begin{abstract}
We present $4.5\ \mu$m luminosity functions for galaxies identified in 178 candidate galaxy clusters at $1.3 < z < 3.2$. The clusters were identified as {\it Spitzer}/IRAC color-selected overdensities in the Clusters Around Radio-Loud AGN (CARLA) project, which imaged 421 powerful radio-loud AGN at $z > 1.3$.  The luminosity functions are derived for different redshift and richness bins, and the IRAC imaging reaches depths of $m^{*} + 2$, allowing us to measure the faint end slopes of the luminosity functions. We find that $\alpha = -1$ describes the luminosity function very well in all redshifts bins and does not evolve significantly. This provides evidence that the rate at which the low mass galaxy population grows through star formation, gets quenched and is replenished by in-falling field galaxies does not have a major net effect on the shape of the luminosity function. Our measurements for $m^{*}$ are consistent with passive evolution models and high formation redshifts ($z_{f} \sim 3$). We find a slight trend towards fainter $m^{*}$ for the richest clusters, implying that the most massive clusters in our sample could contain older stellar populations, yet another example of cosmic downsizing. Modelling shows that a contribution of a star-forming population of up to 40 \% cannot be ruled out. This value, found from our targeted survey, is significantly lower than the values found for slightly lower redshift, $z \sim 1$, clusters found in wide-field surveys. The results are consistent with cosmic downsizing, as the clusters studied here were all found in the vicinity of radio-loud AGNs -- which have proven to be preferentially located in massive dark matter halos in the richest environments at high redshift -- and may therefore be older and more evolved systems than the general protocluster population.  

\end{abstract}


\section{Introduction}

Many attempts have been made to measure the formation epoch of
galaxy clusters, generally finding high formation redshifts, $z_f
\sim 2 - 4$.  Studies focusing on galaxy colors provide insight
into when stellar populations formed \citep[e.g.,][]{Stanford_1998,
Holden_2004, Eisenhardt_2008}, while the assembly of galaxies and
their evolution can be measured by analysing the fundamental plane
or galaxy luminosity functions \citep[e.g.,][]{Van_Dokkum_2003,
Mancone_2010}.

As a few examples, \citet{Eisenhardt_2008} infer stellar formation
redshifts of $z_f > 4$ for cluster galaxies by comparing their
$I-[3.6]$ colors to passive galaxy evolution models. Studying the
color and the scatter of the main sequence, \citet{Mei_2006} infer
a mean luminosity-weighted formation redshift of $z_f > 2.8$ for
cluster ellipticals in two high-redshift clusters in the Lynx
supercluster.  Earlier formation epochs are inferred for galaxies
closer to the cores, with early-type galaxies within 1 arcmin of
the cluster centers having $z_f > 3.7$.  \citet{Kurk_2009}, who
also look at the position of the red sequence in color-magnitude
diagrams and compare them to theoretical predictions, infer $z_f
\sim 3$ for a protocluster at $z = 1.6$.  By comparing the fundamental
plane of a Lynx cluster at $z=1.27$ to the fundamental plane of the
nearby Coma cluster, \citet{Van_Dokkum_2003} infer a stellar formation
redshift of $z_f = 2.6$ for the distant cluster, with passive evolution
thereafter.

Studying the mid-infrared (mid-IR) luminosity function at high
redshifts ($1 < z <3$) probes rest-frame near-infrared (near-IR)
emission ($J,H,K$), which is a good proxy for stellar mass for all
but the youngest starbursting galaxies \citep{Muzzin_2008, Ilbert_2010}.
Such studies have shown that the bulk of the stellar mass in clusters
is already in place by $z \sim 1.3$ \citep[e.g.,][]{Lin_2006,
Muzzin_2008, Mancone_2010} and that $\alpha$, the faint end slope
of the galaxy luminosity function, does not evolve significantly
with redshift \citep[e.g.,][]{Propris_1998, Muzzin_2007, Strazzullo_2010,
Mancone_2012}. It seems that processes that might lead to a substantial
increase in mass such as mergers and star formation, and processes
that would strip mass away from the cluster such as galaxy-galaxy
interactions, or galaxy harassment, either balance each other or
do not to play an important role in cluster evolution.

On the other hand, there is evidence for considerable stochastic
star formation in clusters at $z > 1.3$. For a sample of 16
spectroscopically confirmed galaxy clusters at $1 < z < 1.5$ selected
from the IRAC Shallow Cluster Survey \citep[ISCS;][]{Eisenhardt_2008},
\citet{Brodwin_2013} show that star formation is occurring at all
radii and increases towards the core of the cluster for clusters
at $z > 1.4$. These clusters were identified as 3-D overdensities
in the Bo\"{o}tes Survey \citep{Stanford_2005, Elston_2006,
Eisenhardt_2008} using a photometric redshift probability distribution
and wavelet analysis \citep{Brodwin_2006}. \citet{Brodwin_2013}
observe a rapid truncation of star formation between $z \sim 1.5$
and $z \sim 1$, by which time the cores of the clusters become
mostly quiescent.  Investigating the color and scatter of the red
sequence galaxies, \citet{Snyder_2012} conclude that at $z \sim
1.5$ significant star formation is occurring and that at this
redshift the red sequence in the centers of clusters was rapidly
growing.


In a related analysis also using the Bo\"otes cluster sample,
\citet{Mancone_2010} measured the evolution of the mid-IR luminosity
function for a sample of galaxy clusters spanning $0.3 < z < 2$.
By measuring the luminosity function and the evolution of $m^*$
compared to theoretical passive evolution models, \citet{Mancone_2010}
found $z_f \sim 2.4$ for the low redshift ($z < 1.3$) portion of
their cluster sample. At higher redshift ($1.3 < z < 1.8$) a
significant deviation from the passive models was measured which
could most likely be explained by ongoing mass assembly at those
redshifts. However, the highest redshift bins suffered from small
sample sizes.

This paper aims to continue and complement these previous results
and extend the luminosity function analysis to higher redshift. We
study the evolution of the luminosity function of almost 200 galaxy
cluster candidates at $1.3 < z < 3.2$ discovered through the Clusters
Around Radio Loud AGN, or CARLA, project \citep{Wylezalek_2013}.
The clusters were found in the fields of radio-loud active galactic
nuclei (RLAGN), including both typical unobscured (e.g., type-1)
radio-loud quasars (RLQs), and obscured (e.g., type-2) radio-loud
AGN, also referred to as radio galaxies (RGs). Significant research
stretching back many decades show that RLAGN belong to the most
massive galaxies in the universe \citep{Lilly_1984, Rocca_2004,
Seymour_2007, Targett_2012} and are preferentially located in rich
environments up to the highest redshifts \citep[e.g.,][]{Minkowski_1960,
Stern_2003, Kurk_2004, Galametz_2012, Venemans_2007, Hutchings_2009,
Hatch_2011, Matsuda_2011, Mayo_2012, Husband_2013, Ramos_2013,
Wylezalek_2013}. As described in \citet{Wylezalek_2013}, the selection
of candidate cluster members in the vicinity of the CARLA targets
is based purely on the mid-IR colors of galaxies around a luminous
RLAGN with a spectroscopic redshift. The selection is thus independent
of galaxy age, morphology, or the presence of a red sequence; we
discuss the selection in more detail in \S~6.  This sample allows
us to perform a statistical study of the mid-IR luminosity functions
of a large sample of very high-redshift galaxy clusters, achieving
statistics never before achieved at these early epochs.

The paper is structured as follows: \S2 describes the observations
and the cluster sample used in this work, \S3 describes the fitting
procedure and estimation of the uncertainties while the results are
presented in \S4. In \S5 we explain the robustness tests carried
out and discuss our measurements in \S6. Section~7 summarises the
work. Throughout the paper we assume $H_{0}$ = 70 km s$^{-1}$
Mpc$^{-1}$, $\Omega_{\rm{m}}$ = 0.3, $\Omega_{\Lambda}$ = 0.7. All
magnitudes and colors are expressed in the AB photometric system
unless stated otherwise.
 
\section{Data}
\subsection{Observations and Data Reduction}

As part of the  \textit{Spitzer} snapshot program `Clusters Around Radio-loud AGN' (CARLA), we observed the fields of 209 high-redshift radio galaxies (HzRGs) and 211 RLQs at $1.3 < z < 3.2$.  A description of the sample selection, observation strategy, data reduction, source extraction, determination of completeness limits and initial scientific results are given in \citet{Wylezalek_2013}. 

Briefly, the fields, covering 5.2\arcmin $\times$ 5.2\arcmin\ each, corresponding to a physical size of $\sim 2.5 \times 2.5$ Mpc for the redshift range of the targeted RLAGN, were mapped with the Infrared Array Camera \citep[IRAC; ][]{Fazio_2004} on board the \textit{Spitzer Space Telescope} at 3.6 and 4.5 $\mu$m (referred to as IRAC1 and IRAC2). The total exposure times give similar depths for both channels with the IRAC2 observations having been tailored to  be slightly deeper for our intended science.The HzRG and RLQ samples have been matched with respect to their redshift and radio-luminosity distributions. The L$_{500\rm{MHz}}$-$z$ plane is covered relatively homogeneously in order to be able to study the environments around the AGN as a function of both redshift and radio luminosity. 

The data were reduced using the MOPEX \citep{Makovoz_2005} package and source extraction was performed using SExtractor \citep{Bertin_1996}  in dual image mode with the IRAC2 image serving as the detection image. The CARLA completeness was measured by comparing the CARLA number counts with number counts from the \textit{Spitzer} UKIDSS Ultra Deep Survey (SpUDS, PI: J. Dunlop) survey. The SpUDS survey is a \textit{Spitzer} Cycle 4 legacy program that observed $\sim$ 1 deg$^{2}$ in the UKIDDS UDS field with IRAC and the Multiband Imaging Spectrometer \citep[MIPS; ][]{Rieke_2004} aboard \textit{Spitzer}. SpUDS reaches 5 $\sigma$ depths of $\sim$ 1 $\mu$Jy (mag$ \simeq$ 24). We determine 95\% completeness of CARLA at magnitudes of [3.6] = 22.6 ($= 3.45\ \mu$Jy) and [4.5] = 22.9 ($ =  2.55\ \mu$Jy). 

\subsection{Cluster Sample}

The cluster candidates  studied in this paper were identified as IRAC color-selected galaxy overdensities in the fields of CARLA RLAGN. In order to isolate galaxy cluster candidates we first identify high-redshift sources ($z > 1.3$) by applying the color cut [3.6]-[4.5]$\geqslant$ -0.1 \citep[e.g. ][]{Papovich_2008}. This color selection has proven to be very efficient at identifying high-redshift galaxies independent of their evolutionary stage. A negative $k$-correction, caused by the 1.6 $\mu$m bump that enters the IRAC bands at $z \sim 1$, leads to an almost constant IRAC2 apparent magnitude at $z >1.3$ and a red [3.6]-[4.5] color \citep{Stern_2005, Eisenhardt_2008, Wylezalek_2013}. We apply a counts-in-cells analysis to such color-selected IRAC sources, which we simply refer to as IRAC-selected sources, to identify overdense fields.

\begin{figure}
\centering
\includegraphics[scale = 0.34]{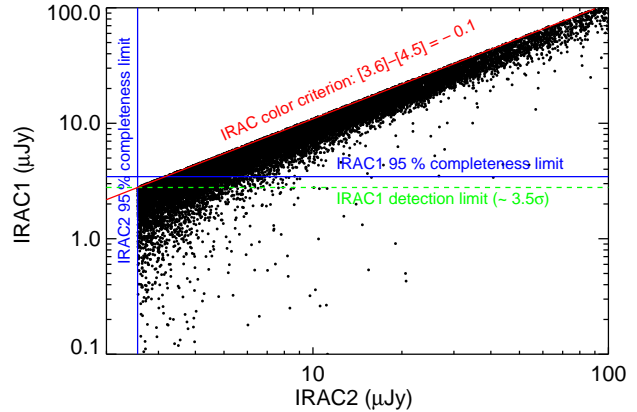}
\caption{IRAC1 flux density versus IRAC2 flux density for all sources in CARLA that pass the color criterion [3.6]-[4.5] $>$ -0.1 (red line). The IRAC1 and IRAC2 95\% completeness limits are shown by the blue horizontal and vertical lines, respectively. The IRAC1 detection limit used in the color criterion analysis (see Section 2.2) is shown by the green dashed line. To build IRAC2 luminosity functions a clear IRAC1 detection is not necessary and we therefore focus on the IRAC2 luminosity function in this paper. The IRAC1 luminosity functions are presented in the Appendix.} 
\label{completeness}
\end{figure}

Specifically, we define a source to be an IRAC-selected source if \textit{(i)} it is detected above the IRAC2 95\% completeness limit and above an IRAC1 flux of 2.8 $\mu$Jy (3.5 $\sigma$ detection limit) and has a color of [3.6]-[4.5] $> - 0.1$ or \textit{(ii)} it is detected above the IRAC2 95\% completeness limit and has an IRAC1 flux $< 2.8\ \mu$Jy but has a color $> -0.1$ at the $3.5 \ \sigma $ detection limit of the IRAC1 observation (e.g. Fig. \ref{completeness}). This refined criterion means that all IRAC-selected sources are 95\% complete in IRAC2 down to [4.5] = 22.9 ($ = 2.55\ \mu$Jy) but are not necessarily well-detected in IRAC1.

We measure the density of IRAC-selected sources in a radius of 1 arcmin centered on the RLAGN and compare it to the mean blank-field density. A radius of 1 arcmin corresponds to $\sim$ 500 kpc over the targeted redshift range and matches typical sizes for $z > 1.3$ mid-IR selected clusters with log$(M_{200}/M_{\sun}) \sim 14.2$ \citep[e.g., ][]{Brodwin_2011}. The typical blank-field density of IRAC-selected sources was measured by placing roughly 500 independent apertures of 1 arcmin radius onto the SpUDS field with its catalogs having been cut at the CARLA depth. Since the publication of \citet{Wylezalek_2013}, 34 new CARLA fields have been observed and we provide the full table of CARLA fields and their overdensities in the electronic journal version of this paper. 

\begin{figure*}
\centering
\includegraphics[scale = 0.6]{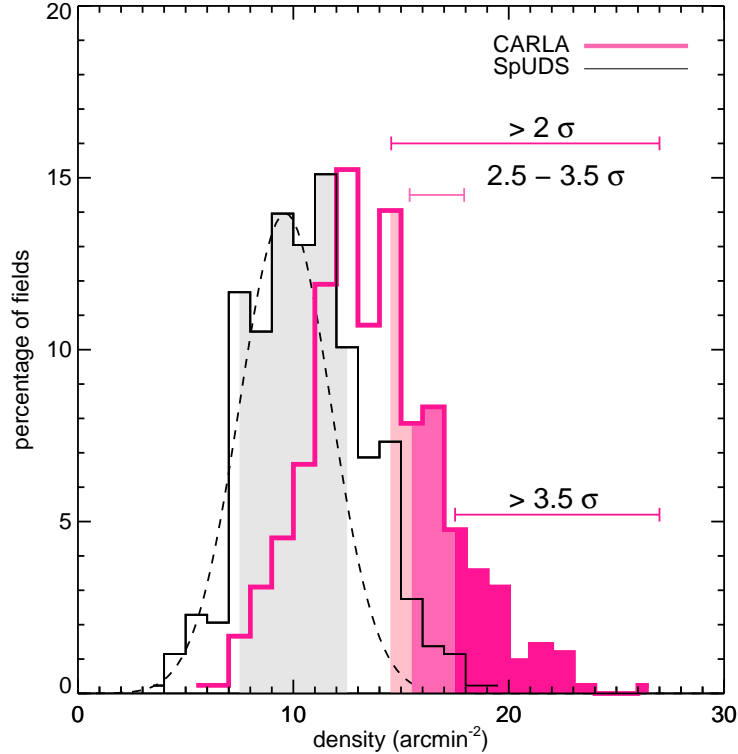}
\caption{Histogram of the surface densities of IRAC-selected sources in the CARLA fields and the SpUDS survey.  Surface densities are measured in circular regions of radius $r = 1$ arcmin. The Gaussian fit to the low-density half of the SpUDS density distribution is shown by the dashed black curve, giving $\Sigma_{\rm{SpUDS}} = 9.6\pm2.1$ arcmin$^{-2}$. The grey shaded area shows all SpUDS cells with a surface density of $\Sigma_{\rm{SpUDS}} = 9.6\pm2.1$ arcmin$^{-2}$, which are used to derive the blank field luminosity function. CARLA clusters are defined as fields with a surface density of $\Sigma_{\rm{CARLA}} > 2 \sigma$. In this paper, however, we also study the dependence of the luminosity function on the CARLA overdensity and repeat the analysis for fields with  $2.5\ \sigma < \Sigma_{\rm{CARLA}} < 3.5\ \sigma$ and  $\Sigma_{\rm{CARLA}} > 3.5\ \sigma$. The fields that go into those analyses are shown by the pink shaded regions, as indicated.}
\label{clusters}
\end{figure*}

In \citet{Wylezalek_2013}, we used the IRAC1 95\% completeness limit of 3.45 $\mu$Jy as the IRAC1 limit in the color selection process, rather than the $3.5\ \sigma$ detection limit of $2.8 \mu$Jy used above. As the IRAC1 95\% completeness limit is brighter than the IRAC2 95\% completeness limit, this IRAC1 limit together with the color criterion introduced an artificially brighter IRAC2 limit  ([4.5] = 22.7, 3.0 $\mu$Jy). In this work we therefore raise the IRAC1 limiting magnitude to [3.6] = 22.8 ($ = 2.8\ \mu$Jy), i.e. we lower the flux density cut, to include all IRAC2 sources down to the formal completeness limit of [4.5] = 22.9. This new IRAC1 flux density cut still corresponds to a $> 3.5 \ \sigma$ detection. This is necessary as we are aiming to include as many faint sources as possible in the analysis. We apply this refined color criterion to the CARLA and SpUDS fields and plot the distribution of their densities in Fig. \ref{clusters}.

Promising galaxy cluster fields are defined as fields that are overdense by at least 2 $\sigma$ \citep[$\Sigma_{\rm{CARLA}} > \Sigma_{\rm{SpUDS}} + 2 \sigma_{\rm{SpUDS}}$, ][]{Wylezalek_2013}. This criterion is met by 46\% of the CARLA fields, 27\% are $\geqslant$ 3 $\sigma$ overdense and 11\% are overdense at the 4 $\sigma$ level. Excluding bad fields (e.g. fields that are contaminated by nearby bright stars), this provides us with 178, 101 and 42 high-redshift galaxy cluster fields at the 2 $\sigma$, 3 $\sigma$ and 4 $\sigma$ level, respectively. 

Note that this selection is a measure of the overdensity signal compared to a blank field, i.e. to a distribution of cell densities centred on random positions. Since the sample is selected using specific galaxies, RLAGN, there will always be at least one source in the overdensity. As all galaxies are clustered to some extent \citep{Coil_2013} the exact measurement of the overdensity signal of RLAGN would be to compare it to a distribution of background cells centred on random galaxies, not random positions. 

However, the average number density of IRAC-selected sources in SpUDS is very high, on average 34 sources per aperture with radius $r = 1$ arcmin. This means that if the IRAC-selected sources were extended and filled the aperture maximally, their radius would only be $r_{ext} =10.3$ arcsec and the average distance between two sources would be $2\times r_{ext} = 20.6$ arcsec. We repeated the blank-field analysis by measuring the densities in roughly 500 apertures centred on random SpUDS IRAC-selected sources. Due to the large aperture radius of $r = 1$ arcmin and small $r_{ext}$, the difference between the two background measurements is not significant ($\Sigma_{\rm{SpUDS}}  = 10.3 \pm 2.6$ arcmin$^{-2}$ compared to $\Sigma_{\rm{SpUDS}}  = 9.6 \pm 2.1$ arcmin$^{-2}$, see Section 3.2). As we will work with the blank field background for the LF analysis and to be consistent with \citet{Wylezalek_2013} we show the SpUDS blank field distribution in Fig. \ref{clusters}.

\citet{Wylezalek_2013} shows that the radial density distribution of the IRAC-selected sources is centered on the RLAGN, implying that the excess IRAC-selected sources are associated with the RLAGN. For the following analysis we assign the redshift of the targeted RLAGN to the IRAC-selected sources in the cell and study the evolution of these galaxy cluster member candidates as a function of redshift. To exclude problematic cluster candidates, we also checked that the median [3.6]$-$[4.5] color of the IRAC-selected sources per field is in agreement with the [3.6]$-$[4.5] color expected for a source with the redshift of the targeted RLAGN. No obvious problematic fields were found. For the rest of the manuscript we will use the term `galaxy clusters' to refer to these cluster and protocluster candidates. 

\section{The Luminosity Function for Galaxy Clusters at $z > 1.3$}
\subsection{Method}

\begin{figure*}
\centering
\includegraphics[scale = 0.4]{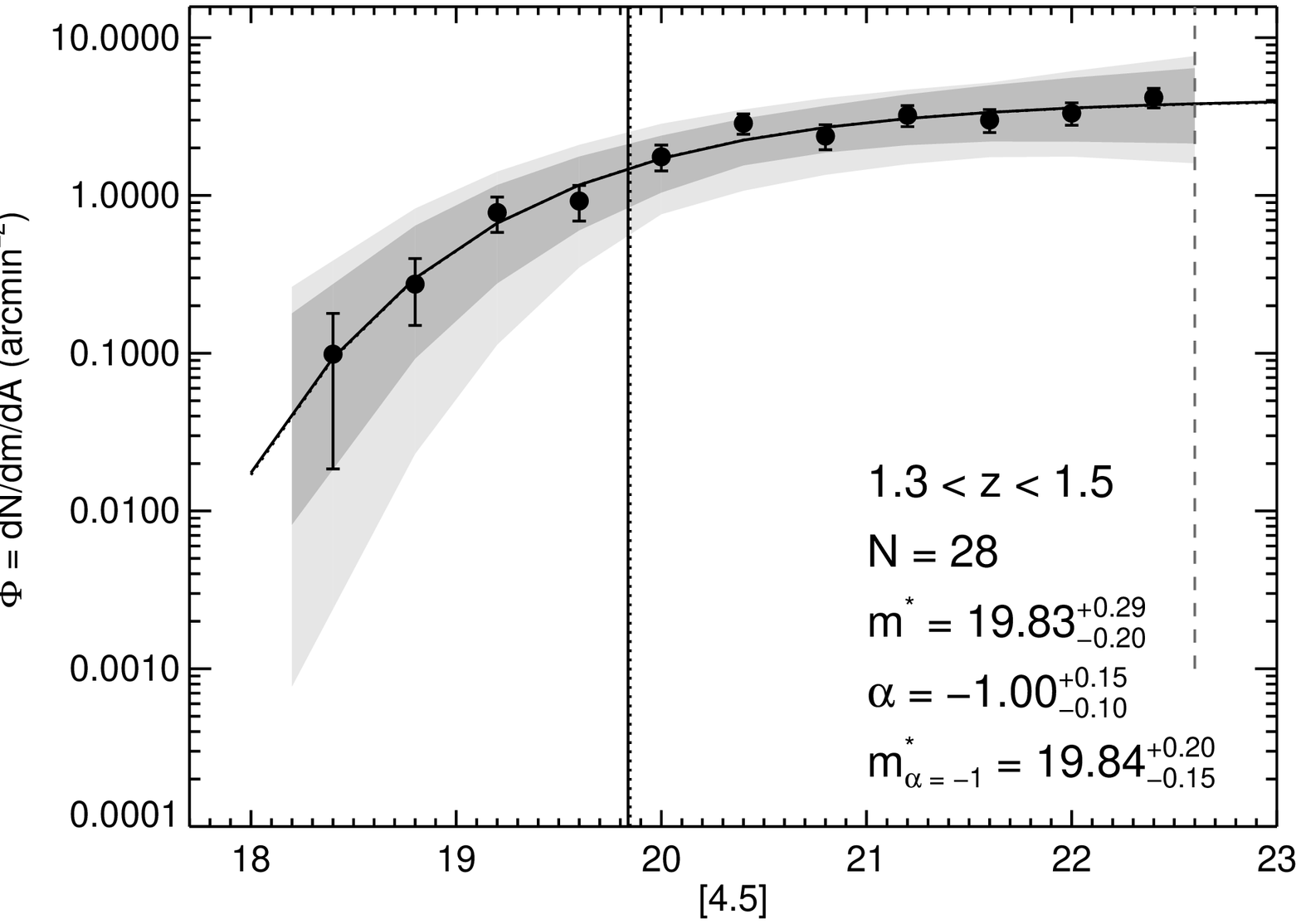}
\includegraphics[scale = 0.4]{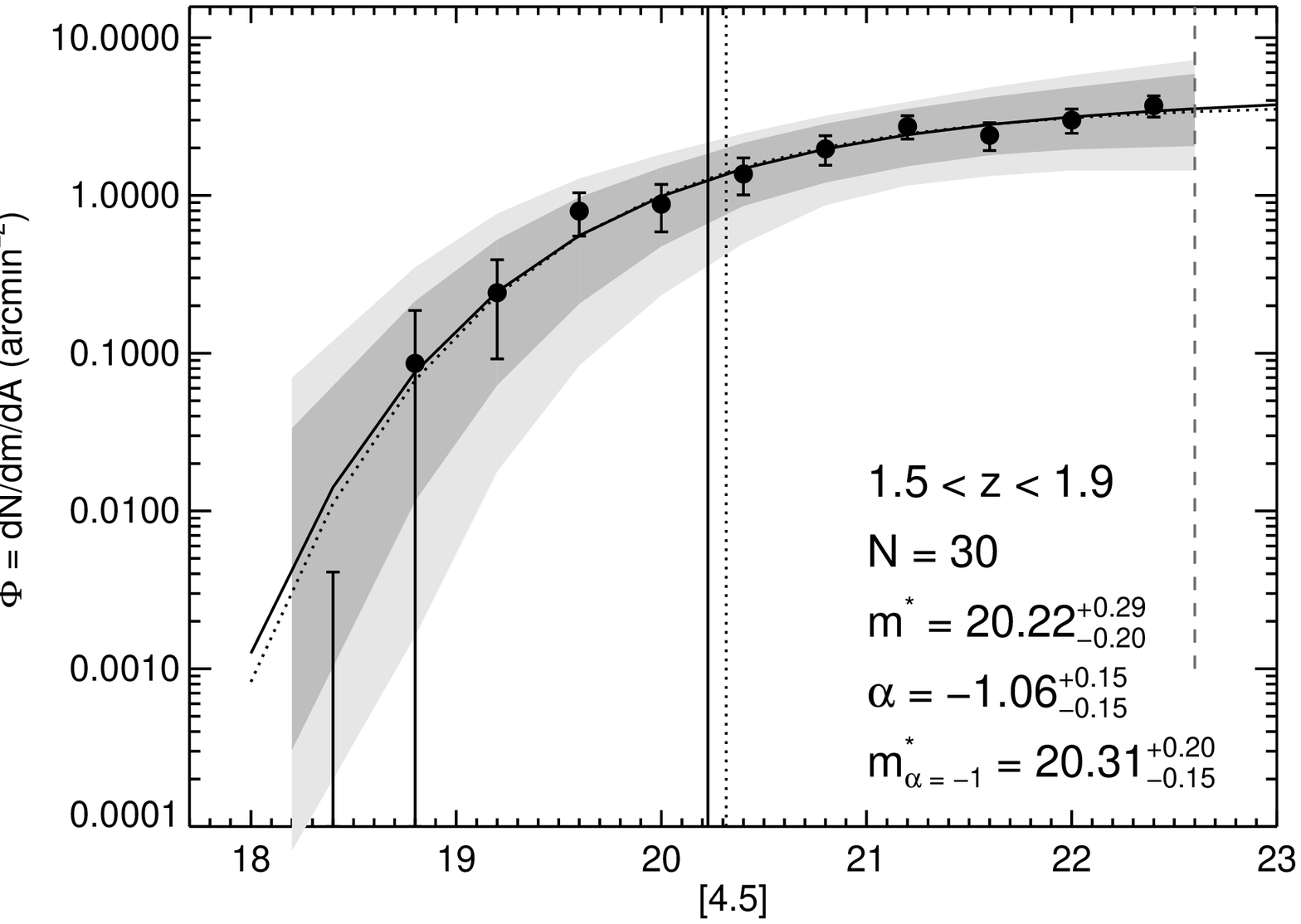}

\includegraphics[scale = 0.4]{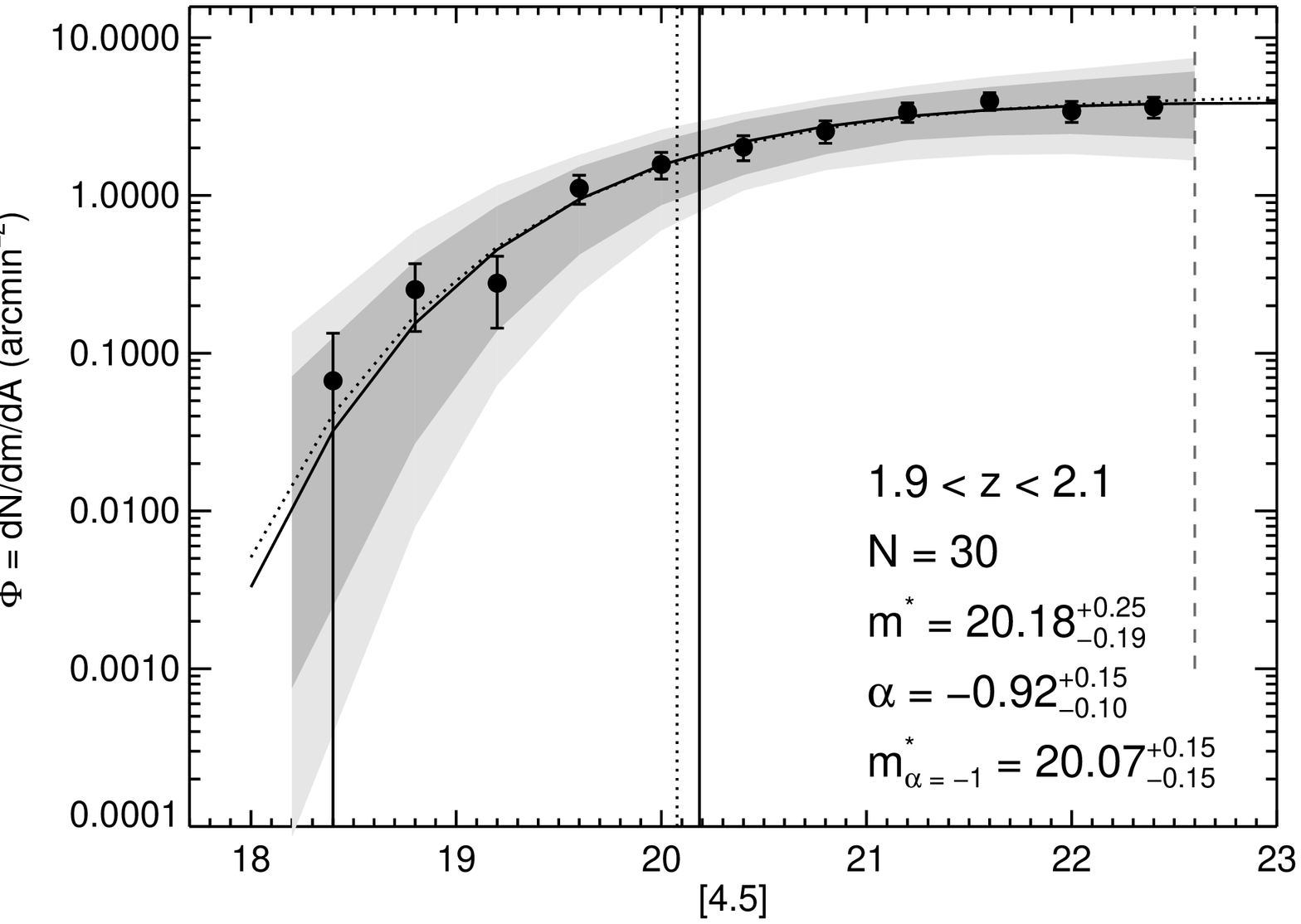}
\includegraphics[scale = 0.4]{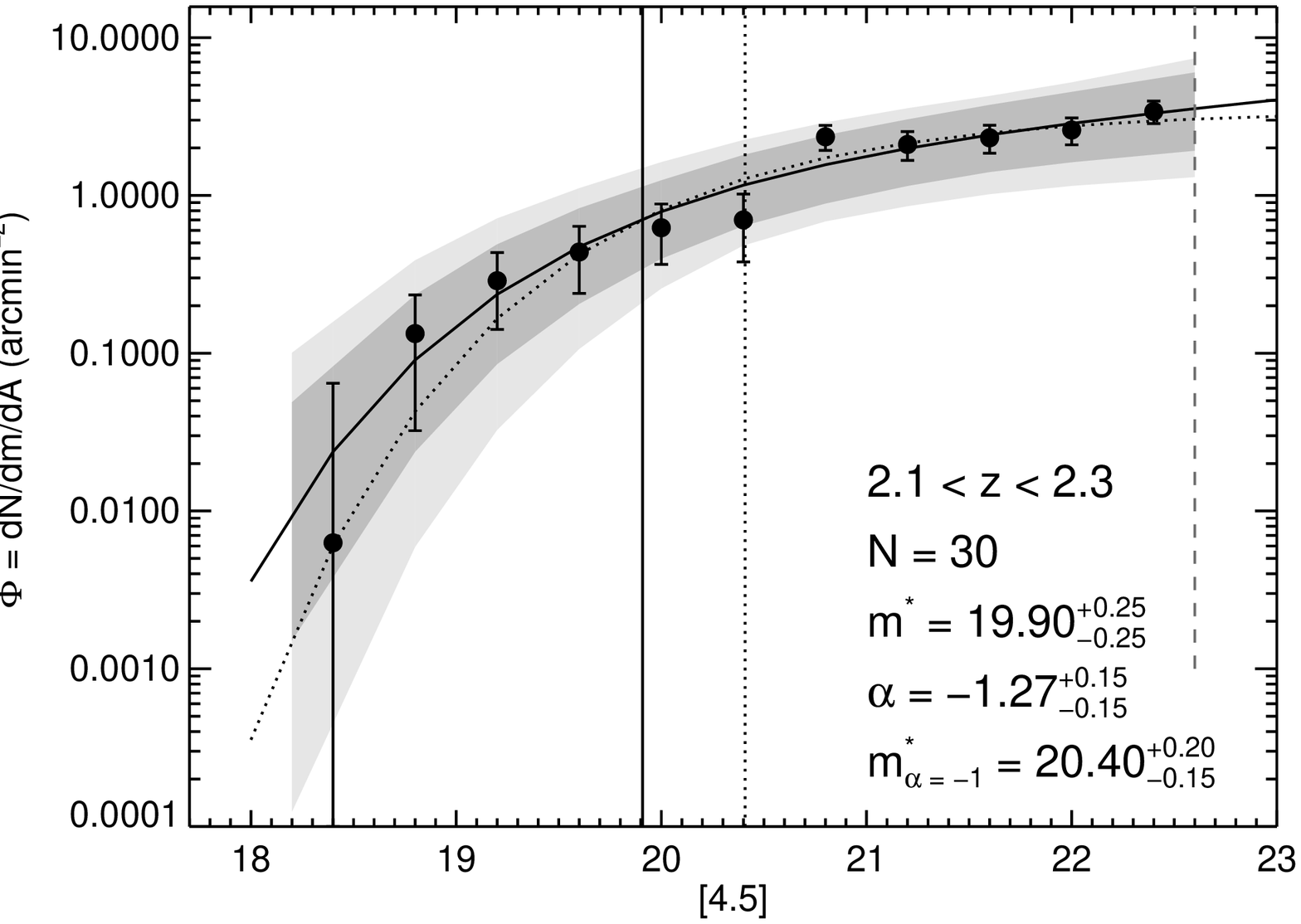}

\includegraphics[scale = 0.4]{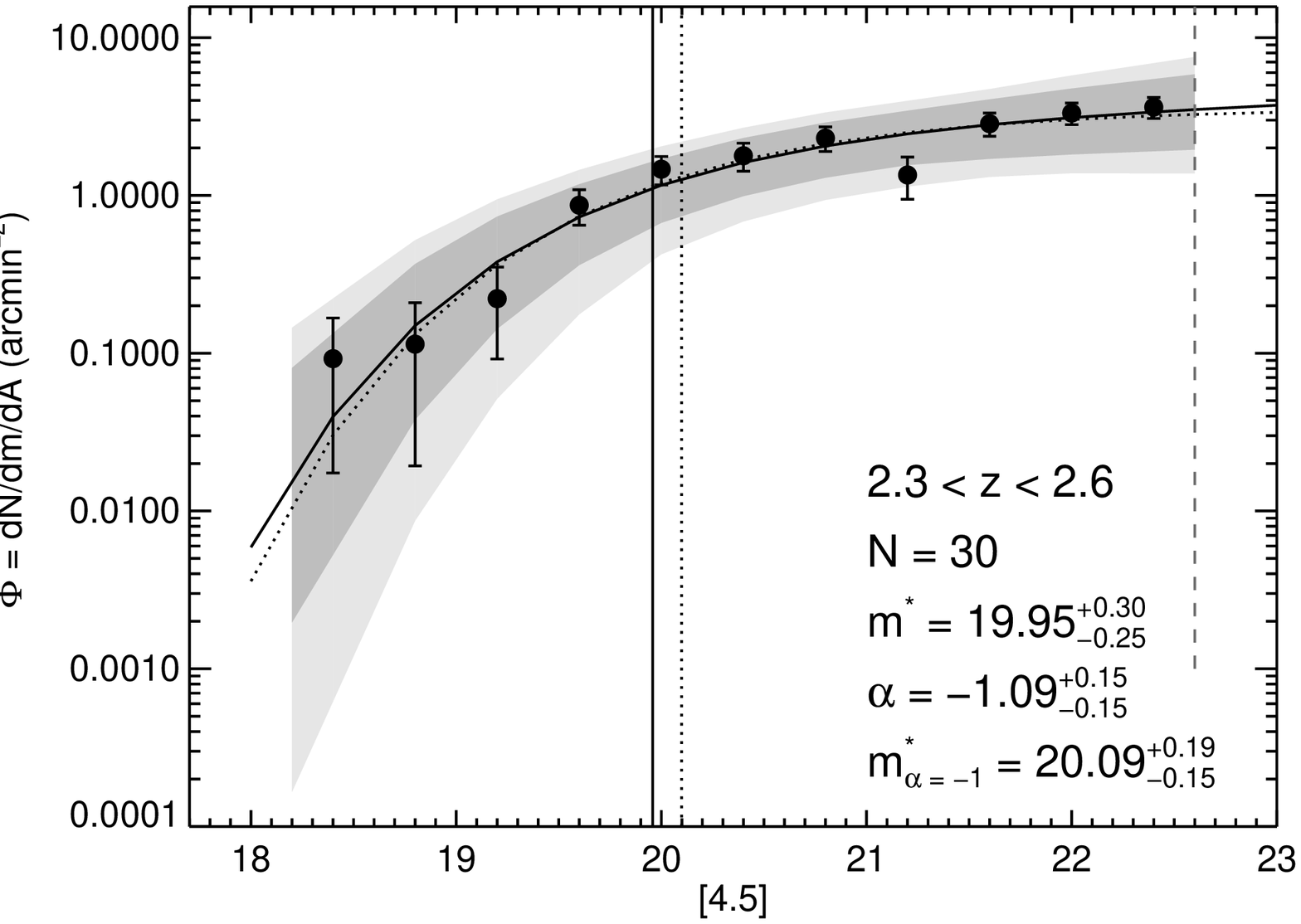}
\includegraphics[scale = 0.4]{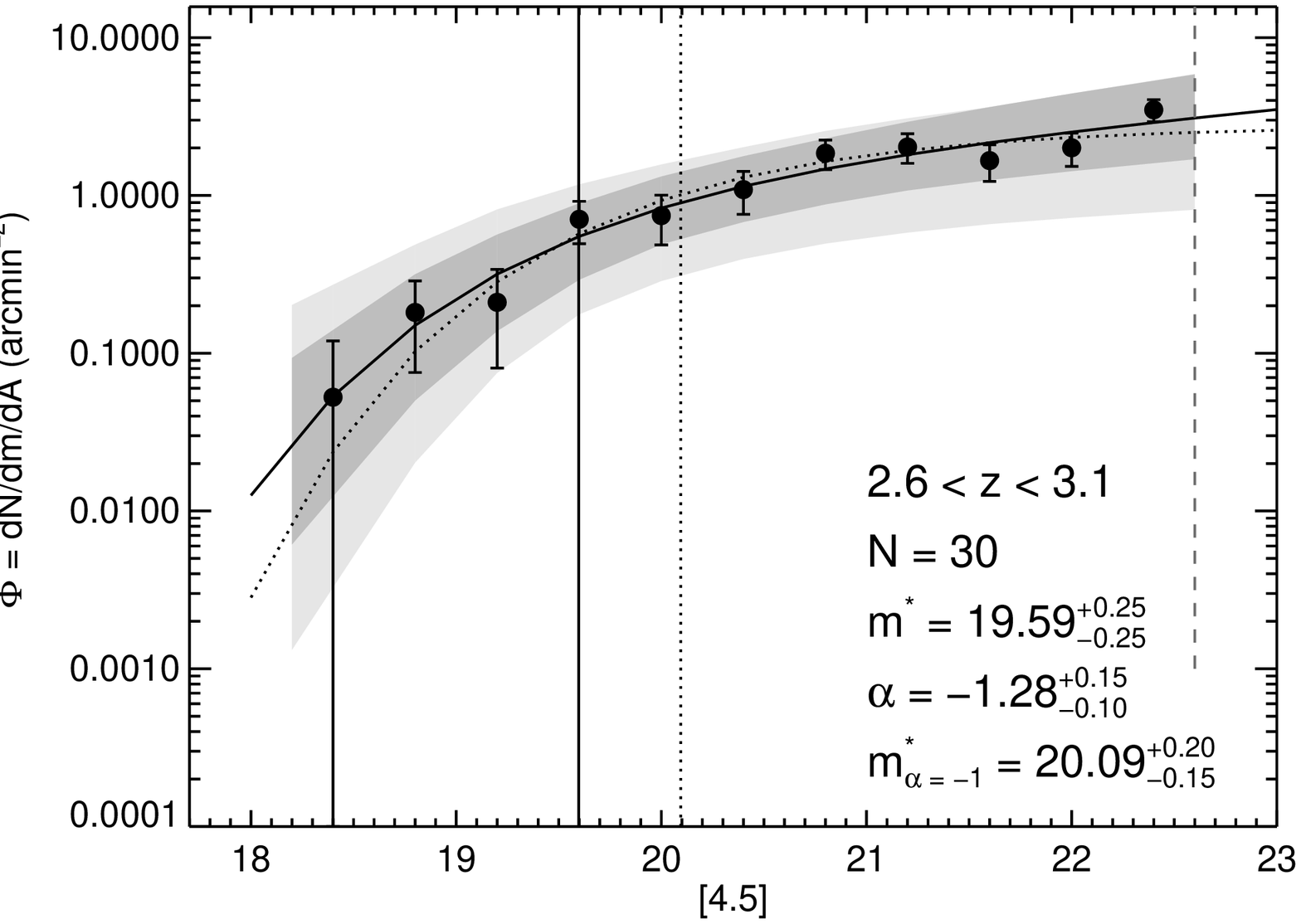}

\caption{Schechter function fits to the 4.5 $\mu$m cluster luminosity function in each redshift bin for CARLA cluster members with $\Sigma_{\rm{CARLA}} > 2 \sigma$. The redshift bins were chosen to contain similar numbers of objects, $N$. The solid circles are the binned differences between the luminosity function for all IRAC-selected sources in the clusters and the background luminosity function derived from the SpUDS survey. The solid curve shows the fit to the data for a free $\alpha$ while the dotted curve shows the fit for $\alpha = -1$. The vertical solid and dotted lines show the fitted values for $m^{*}$ for free and fixed $\alpha$, respectively. The dark and light grey shaded regions show the $1 \sigma$ and $2 \sigma$ confidence regions for the $\alpha$-free-fit derived from Markov Chain Monte Carlo simulations. The vertical dashed line shows the apparent magnitude limit. Note that due to binning of the data and the application of $k$- and $e$-corrections, that are largest for the low-redshift LFs, we here show the 99\% magnitude limit.}
\label{sample}
\end{figure*}

Luminosity functions, $\Phi(L)$, provide a powerful tool to study the distribution of galaxies over cosmological time. They measure the comoving number density of galaxies per luminosity bin, such that
\begin{equation}
dN = \Phi(L)dLdV
\end{equation}
where $dN$ is the number of observed galaxies in volume $dV$ within the luminosity range $[L, L+dL]$. 

There are many ways in which $\Phi(L)$ can be estimated and parametrised, but the most common of these models is the Schechter function \citep{Schechter_1976} 
\begin{equation}
\Phi(L)dL = \Phi^{*}(\frac{L}{L^{*}})^{\alpha}\exp(\frac{-L}{L^{*}})\frac{dL}{L^{*}}
\end{equation}
where $\Phi^{*}$ is a normalisation factor defining the overall density of galaxies, usually quoted in units of $h^{3}$Mpc$^{-3}$, and $L^{*}$ is the characteristic luminosity. The quantity $\alpha$ defines the faint-end slope of the luminosity function and is typically negative, implying large numbers of galaxies with faint luminosities. The luminosity function can be converted from absolute luminosities $L$ to apparent magnitudes $m$ and be written as
\begin{equation}
\label{eq:LF}
\Phi(m)= 0.4\ln(10)\Phi^{*}\frac{(10^{0.4(m^{*}-m)})^{(\alpha+1)}}{e^{10^{0.4(m^{*}-m)}}}
\end{equation}
where $m^{*}$ is the characteristic magnitude. 

In this work we study the evolution of $m^{*}$ in galaxy clusters as a function of redshift for both $\alpha$ as a free fitting parameter and fixed $\alpha =-1$. We include all CARLA fields that are overdense at the 2 $\sigma$ level or more. 

We measure the IRAC2 luminosity function of the galaxy clusters from the CARLA survey as a function of redshift, defined as the redshift of the AGN, and galaxy cluster richness. The CARLA galaxy cluster richness is defined in terms of the significance of the overdensity of IRAC-selected sources within the cell centered on the RLAGN (Fig. \ref{clusters}). For the largest sample of CARLA clusters, i.e. those overdense at the  $ > 2 \ \sigma$ level, we measure the luminosity function in six redshift bins chosen in a way that all redshift bins contain the same number of galaxy clusters. We also consider CARLA clusters overdense at the $2.5 -3.5 \ \sigma$ and  $> 3.5\ \sigma$ level. For these smaller subsamples we measure the luminosity function in three redshift bins. IRAC1 luminosity functions are presented in the Appendix.

For each CARLA cluster candidate $j$ we compute the $k$- and evolutionary corrections ($k_j$- and $e_j$) that are required to shift the galaxy cluster members to the center of the redshift bin and apply them to the apparent magnitudes of the galaxy cluster members. The corrections were computed using the publicly available model calculator EzGal \citep{Manconeez_2012} using the predictions of passively evolving stellar populations from \citet{Bruzual_2003} with a single exponentially decaying burst of star formation with $\tau = 0.1$ Gyr and a Salpeter initial mass function. The $k_j$+$e_j$ corrections are of order 0.04 mag and do not have a significant impact on the LF analysis; they are simply applied for completeness. The typically bright targeted radio-loud AGNs have been removed from the analysis. 

For each CARLA cluster candidate $j$ we measure the number of cluster members $ n_{m_{i}}$ (in a cell with radius $r = 1$ arcmin centered on the RLAGN) in the $i^{th}$ magnitude bin with $m_{i} = [m_i,m_i+\delta m_i]$:

\begin{equation}
\Phi_{i_{\rm{CARLA}}, j, k_j+e_j } = n_{m_{i}}
\end{equation}
This number density is a superposition of the cluster luminosity function and the luminosity function of background/foreground galaxies. In the following we describe the statistical background determination and how the true galaxy cluster luminosity function is determined.

\subsection{Background Subtraction}

The CARLA fields cover an area of $\sim$ 5.2\arcmin $\times$ 5.2\arcmin, roughly corresponding to a region with a radius of 1--1.5 Mpc for the typical redshift of the RLAGN. This radius is in good agreement with sizes of typical mid-IR selected clusters \citep[e.g.][]{Brodwin_2011}. Therefore a local background subtraction in each field is not possible. Instead, we determine a global background in a statistical way using the SpUDS survey. 

As described in Section 2.2, we placed roughly 500 random, independent (i.e. non-overlapping) apertures with radius $r = 1$ arcmin onto the SpUDS survey to estimate the typical blank field density of IRAC-selected sources. Fitting a Gaussian to the low-density half of the SpUDS density distribution finds a mean surface density of $\Sigma_{\rm SpUDS} = 9.6$ arcmin$^{-2}$ with a width of $\sigma_{\rm SpUDS} = 2.1$ arcmin$^{-2}$.  The tail at larger densities arises as even a 1~deg$^2$ survey has large scale structure and contains clusters. To determine the mean background, we consider cells in the SpUDS survey with surface densities of IRAC-selected sources in the range $9.6 \pm 2.1$ arcmin$^{-2}$ (see Fig. \ref{clusters}).

The average background luminosity function per cell is given by

\begin{equation}
\Phi_{i_{\rm{BG}}} = n_{m_{i}, \rm{BG}}/N_{\rm{BG}}
\end{equation} 
where $n_{m_{i}, \rm{BG}}$ is the number of galaxies with magnitudes $m_{i} = [m_i,m_i+\delta m_i]$ in the SpUDS background cells and $N_{\rm{BG}}$ is the number of SpUDS cells used for the background determination. 

We then subtract this average blank field luminosity function from the luminosity function of each CARLA cluster field $j$ after having applied the same $k_j-$ and $e_j-$correction as for the corresponding CARLA field, such that

\begin{equation}
\Phi_{i,j} = \Phi_{i_{\rm{CARLA}},j, k_j+e_j } - \Phi_{i_{\rm{BG}}, k_{j}+e_{j} } 
\end{equation}
After this background subtraction the signal of all CARLA cluster fields per redshift bin is stacked to obtain the background-subtracted luminosity function:

\begin{equation}
\Phi_{i} = (\sum\limits_{j=0}^{N}{\Phi_{i,j}} )\times \frac{1}{N}\times\frac{1}{A}
\end{equation}
where $N$ is the number of CARLA clusters in the redshift bin and $A = 1\ \rm{arcmin}^{2}\times\pi$ is the area of a cell with a radius of 1 arcmin. Figure \ref{sample} shows the luminosity functions for CARLA fields of richness $\geq 2\sigma$ divided into six redshift bins.

The accurate measurement of the background number counts is essential for measuring the true cluster luminosity function. Because the SpUDS survey, used in this paper, covers only $\sim$ 1 deg$^{2}$, it needs to be confirmed that it represents a typical blank field and is not significantly affected by cosmic variance. At the depth of the CARLA observations,however, it is the largest contiguous survey accessible. 

\begin{figure}
\centering
\includegraphics[scale = 0.45]{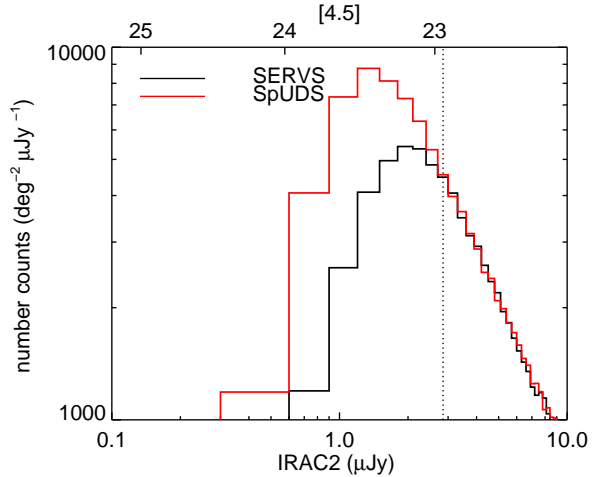}  
\caption{Number counts for the SpUDS and SERVS survey for sources detected at 4.5 $\mu$m. The 95 \% completeness limit of SERVS is derived by comparing the SERVS number counts to those from the SpUDS survey, which gives a completeness limit of 2.85 $\mu$Jy for IRAC2, as shown by the vertical dotted line.}
\label{spuds_servs}
\end{figure}

The 18 deg$^2$ \textit{Spitzer} Extragalactic Representative Volume Survey \citep[SERVS, ][]{Mauduit_2012} reaches almost the same depth as CARLA and allows a test of the goodness of SpUDS as a blank field. SERVS maps five well observed astronomical fields (ELAIS-N1, ELAIS-S1, Lockman Hole, \textit{Chandra} Deep Field South and \textit{XMM}-LSS) with IRAC1 and IRAC2. Coverage is not completely uniform across the fields but averages $\sim$ 1400 s of exposure time. We extracted sources from the SERVS images (Mark Lacy, priv. communication) in the same way as we did for CARLA and SpUDS to allow for a consistent comparison. In Fig. \ref{spuds_servs} we show the IRAC2 number counts in SpUDS and SERVS observations of the \textit{XMM}-LSS field, illustrating the difference in depths of the two surveys. Comparing the SERVS number counts with the SpUDS number counts gives an IRAC2 95 \% completeness limit of 2.85 $\mu$Jy for the SERVS observations. As we aim to go as deep as possible for the CARLA analysis, the SERVS 95\% completeness is slightly shallow compared to the corresponding depth of 2.55 $\mu$Jy
for the CARLA observations.  We therefore use the smaller area SpUDS survey for the background determinations.

\begin{figure}
\centering
\includegraphics[scale = 0.31]{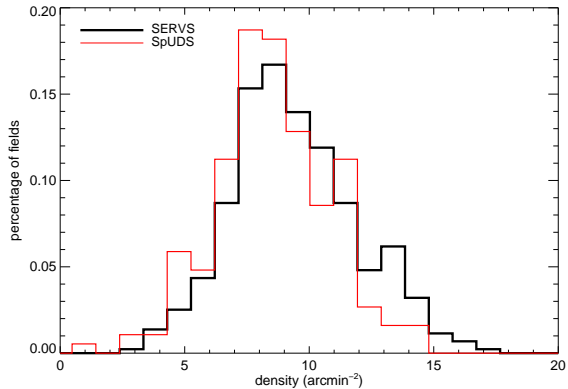}
\caption{Distribution of IRAC-selected sources in the SpUDS and SERVS surveys. The apertures placed on SpUDS cover almost the full area of SpUDS and will pick up any cluster or large scale structure in that field. By placing apertures onto the much wider SERVS survey that is less biased by large-scale structure the high-density tail becomes less prominent, as expected. This motivates using SpUDS cells with $\Sigma_{\rm{SpUDS}}\pm\sigma_{\rm{SpUDS}}$ as a sensible measure of the blank field density of IRAC-selected sources.}
\label{histo_servs}
\end{figure}

To test the validity of our background subtraction we placed 200 random apertures onto the 18 deg$^{2}$ SERVS survey and measured the density of IRAC-selected sources in those random cells. There is a chance that a CARLA cluster field and a non-associated large scale structure are found in projection. In this case our background subtraction would underestimate the actual background in that field. By placing 200 random apertures onto the 18 deg$^{2}$ SERVS survey we can get a qualitative upper limit of this probability. Figure \ref{histo_servs} shows the distribution of densities of IRAC-selected sources in those 200 random cells at the magnitude limit of the SERVS survey and compares it to the distribution in SpUDS. SpUDS only covers $\sim$ 1 deg$^{2}$ and is known to be biased to contain clusters and large scale structure. The apertures placed on SpUDS cover almost the full area and will thus pick up any large scale structure or clusters in that field. That is why a prominent high density tail arises. By placing random apertures onto a larger survey that is less biased by large-scale structure the high-density tail becomes less prominent, as expected. 

\begin{figure}
\centering
\includegraphics[scale = 0.4]{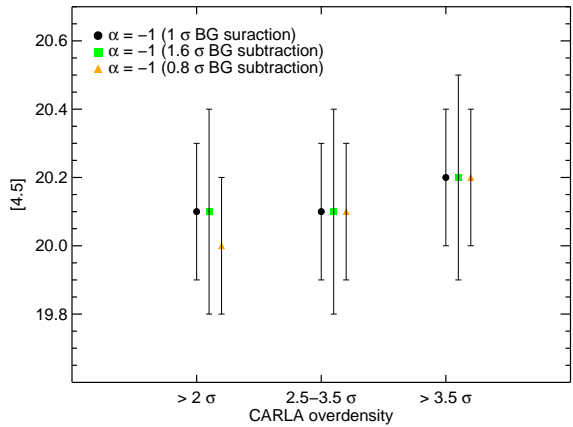}  
\caption{Median best-fit $m^*_{4.5}$ with $\alpha = -1$ as a function of galaxy cluster candidate richness for three different background-subtraction intervals. This demonstrates that the exact prescription of the background interval has an insignificant effect on the derived parameters.  For clarity, data points for each background interval are shifted slightly along the horizontal axis.}
\label{density_evol}
\end{figure}

To make sure, however, that we do not underestimate the background luminosity function by not taking enough of the high density tail into account, we test here what impact our choice of background has on the overall results in this work. We repeat the luminosity function fitting analysis by choosing different density intervals to estimate the background, namely $\Sigma_{\rm{SpUDS}}\pm0.8 \times \sigma_{\rm{SpUDS}}$ and $\Sigma_{\rm{SpUDS}}\pm1.6 \times \sigma_{\rm{SpUDS}}$. Taking an even narrower interval than $\Sigma_{\rm{SpUDS}} \pm 0.8 \times \sigma_{\rm{SpUDS}}$ gives too few apertures and results in a very noisy background luminosity function. In Fig. \ref{density_evol} we show the results for the luminosity function fits with fixed $\alpha$ for the original background subtraction ($1\ \sigma$) and for the test background subtractions ($0.8\ \sigma, 1.6\ \sigma$). The results for the different runs agree remarkably well and no systematic effect is seen. At  $\Sigma_{\rm{SpUDS}} = \Sigma_{\rm{SpUDS}}\pm1.6 \times \sigma_{\rm{SpUDS}}$ we are already sampling part of the high density tail. Table \ref{tablecarla} shows that this has a minimal impact on the Schechter function fits. The exact choice of the density interval of the background subtraction is not critical and illustrates that choosing SpUDS cells with $\Sigma_{\rm{SpUDS}}\pm\sigma_{\rm{SpUDS}}$ is a sensible measure for the blank field density of IRAC-selected sources.

\begin{figure*}
\centering
\includegraphics[scale = 0.43]{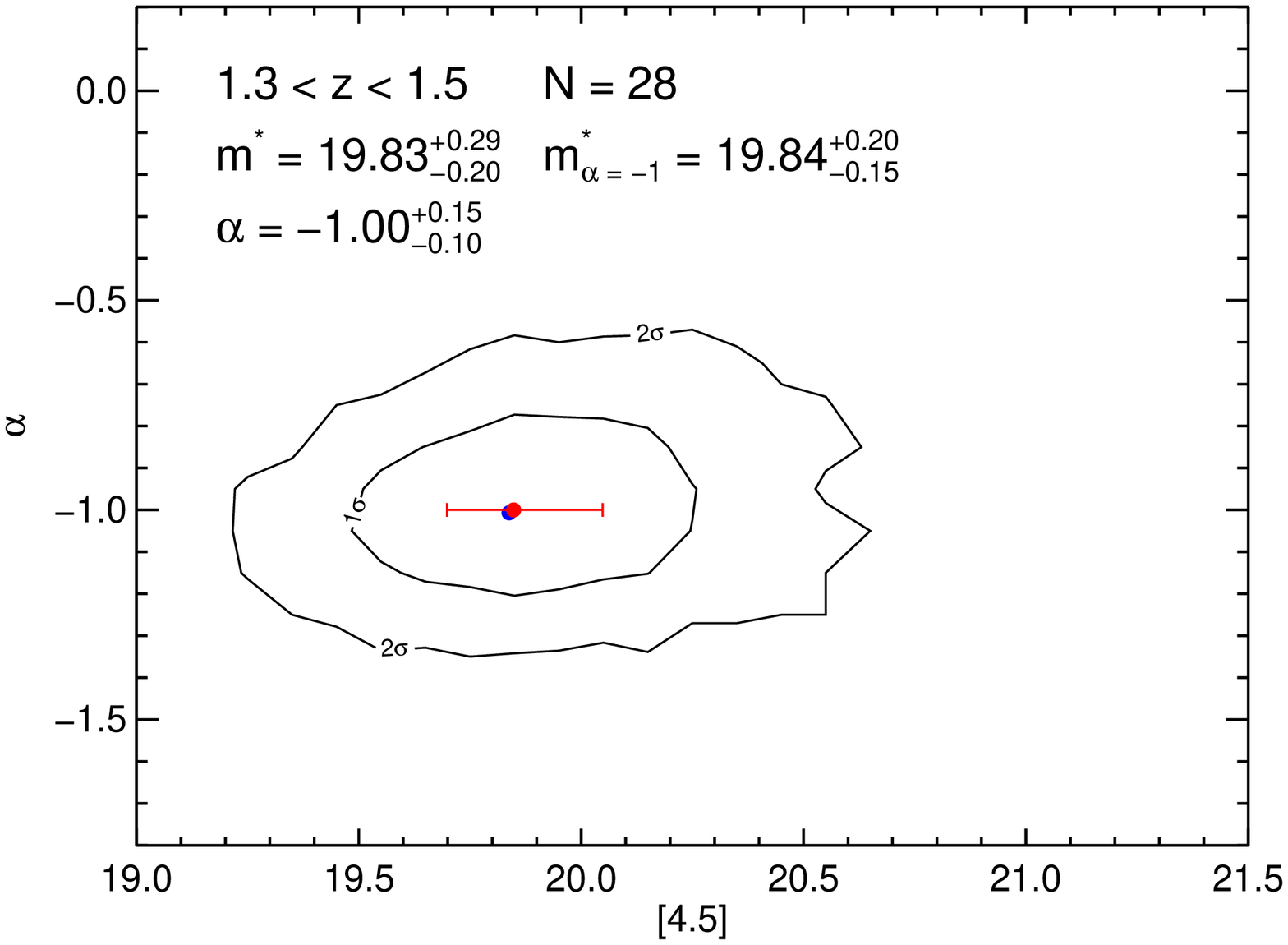}  
\includegraphics[scale = 0.43]{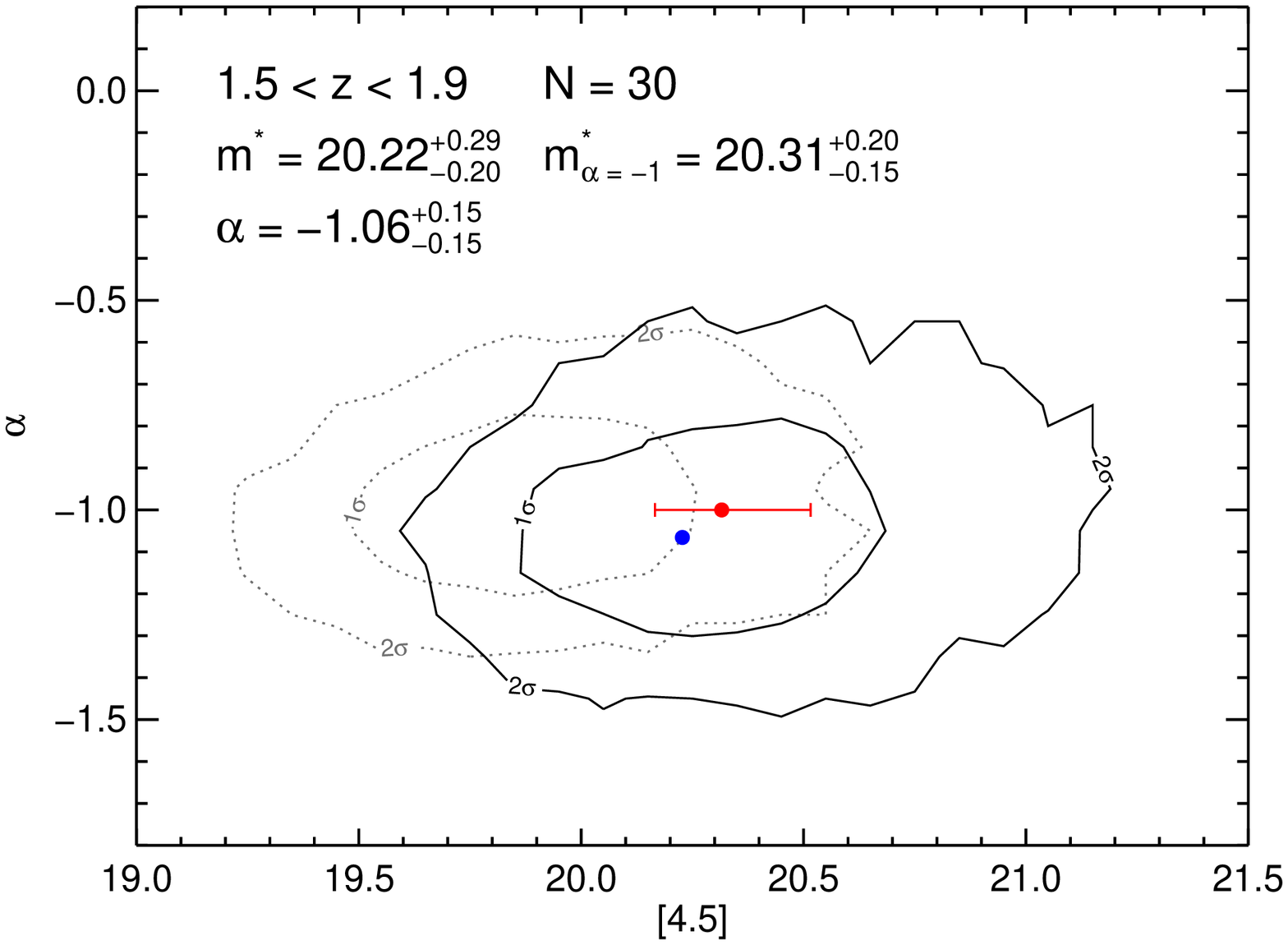}

\includegraphics[scale = 0.43]{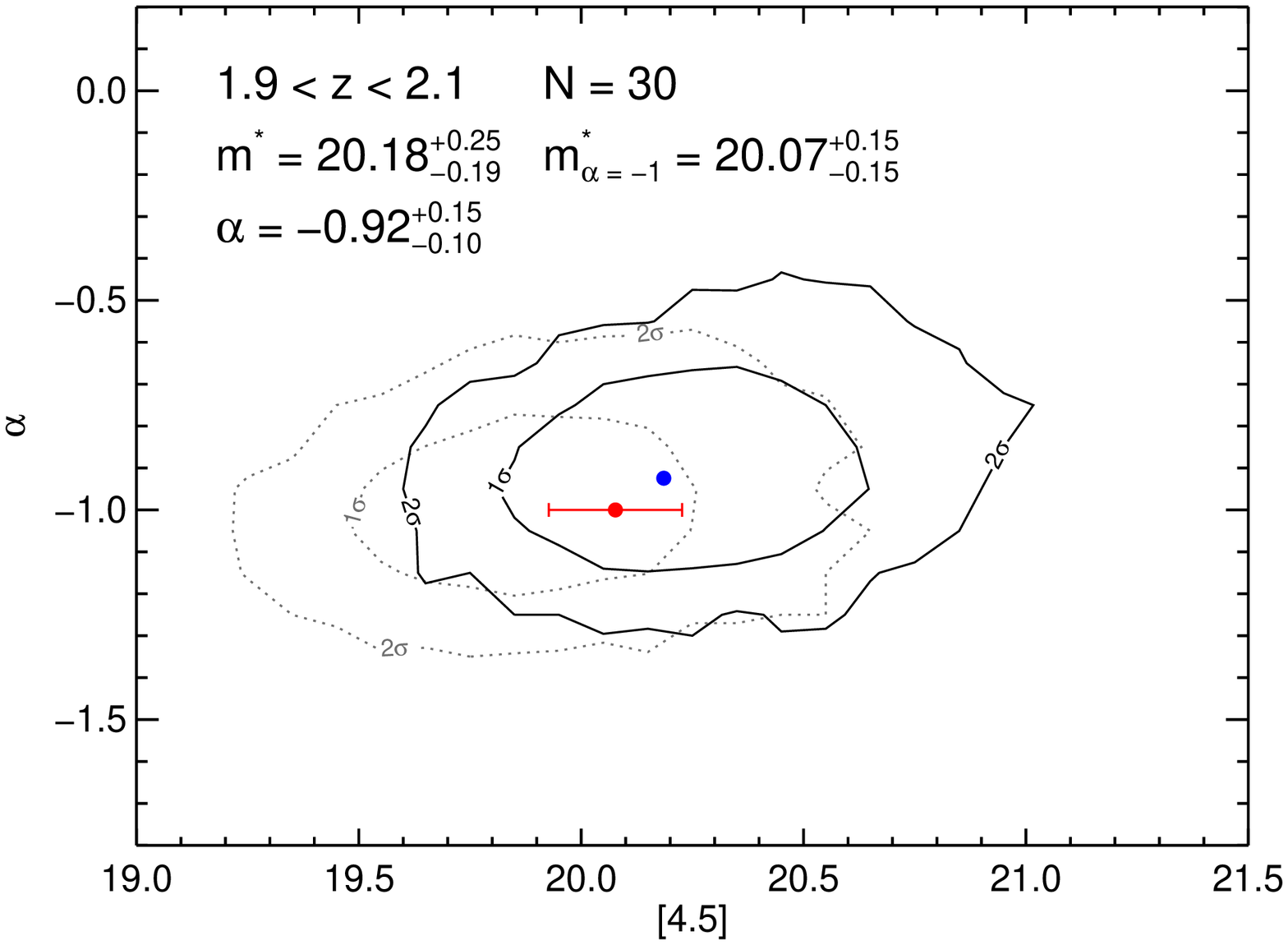}
\includegraphics[scale = 0.43]{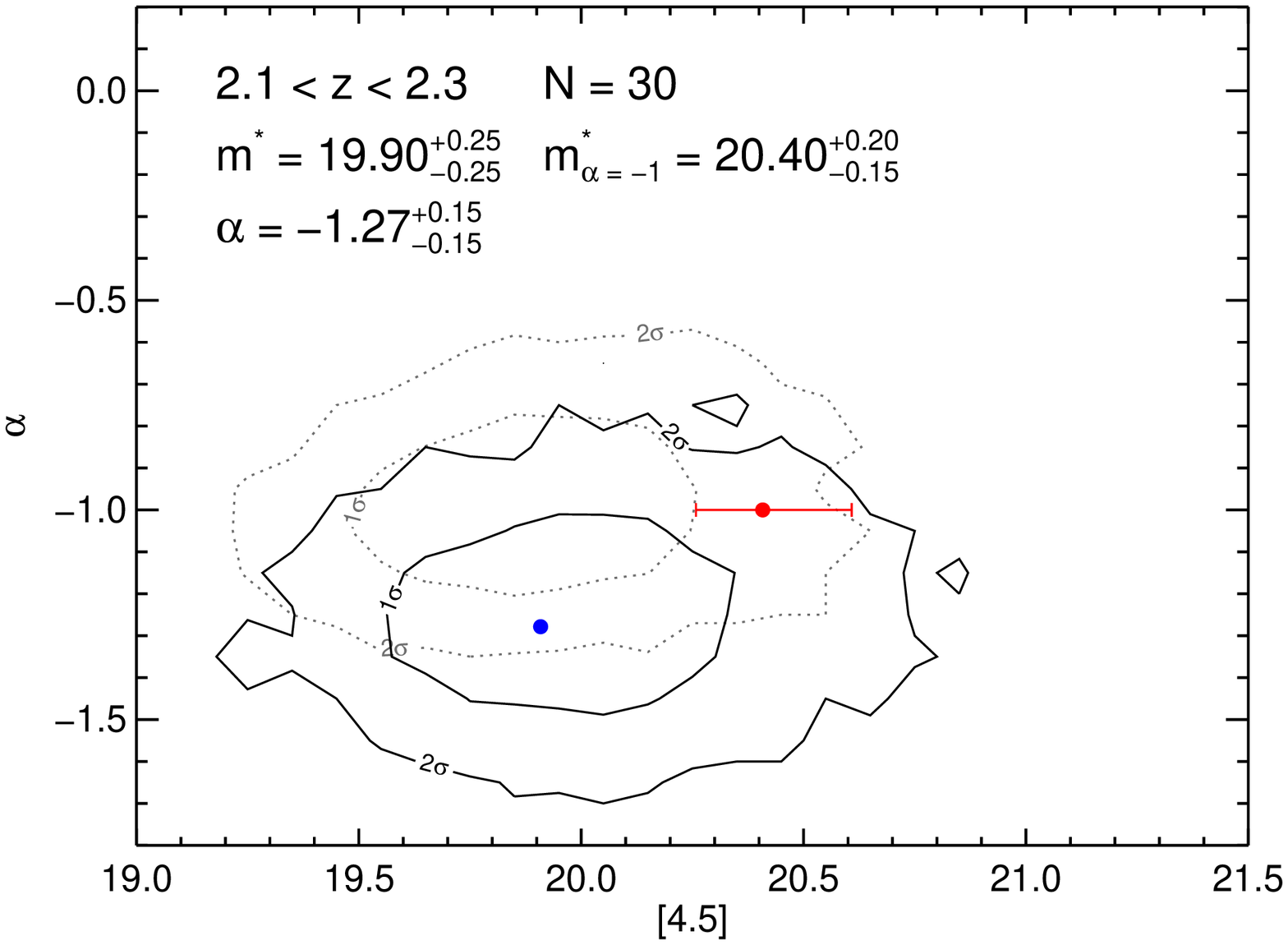}

\includegraphics[scale = 0.43]{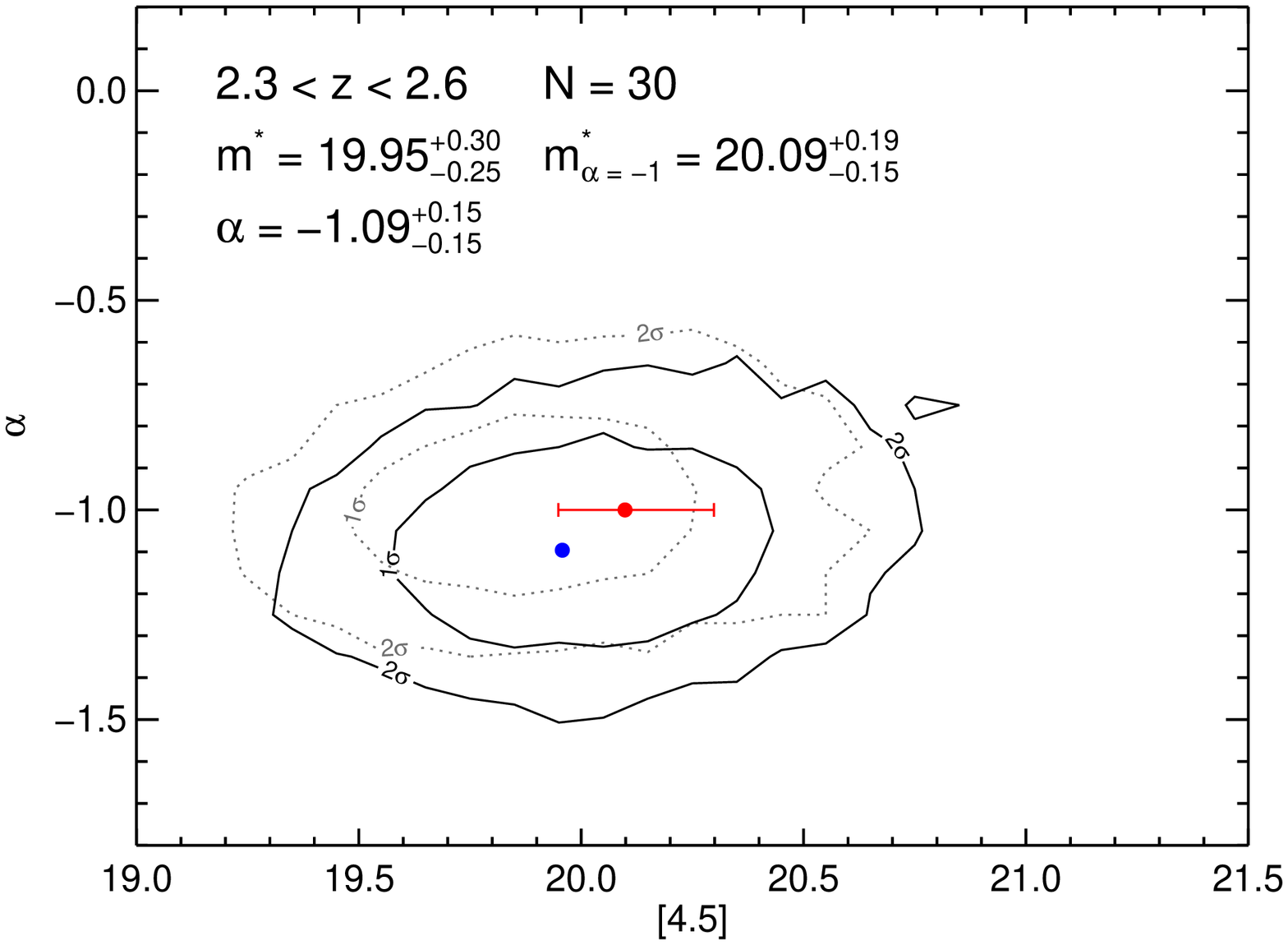}
\includegraphics[scale = 0.43]{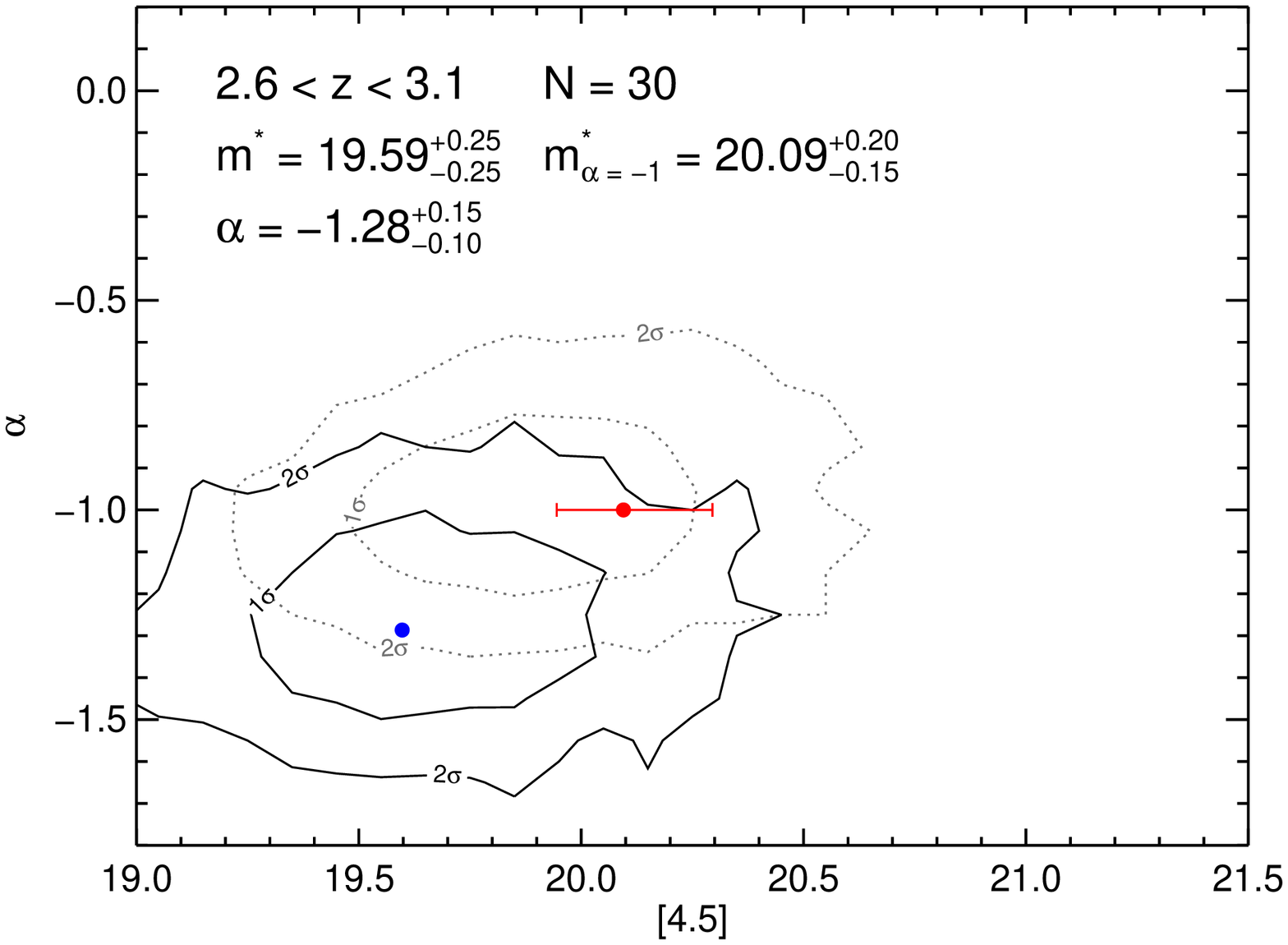}

\caption{Confidence regions for $\alpha$ vs. m$^*$ for Schechter fits with $\alpha$ as a free parameter derived from the Markov Chain Monte Carlo simulations. The contours show the 1 $\sigma$ and 2 $\sigma$ contour levels and the blue solid circle shows the result of the best Schechter fit using the Levenberg-Marquart technique. The redshift bins were chosen to contain similar numbers of objects, $N$. The red solid circle with uncertainty on $m^*$ shows the best Schechter fit for a fixed $\alpha=-1$. In all cases the results from the fixed $\alpha$ fit and the free $\alpha$ fit agree within their confidence regions implying that $\alpha = -1$ describes the luminosity function well over the whole redshift range probed in this work. For comparison, we show the lowest redshift contours (dotted grey) in the higher redshift panels.}
\label{mcmc}
\end{figure*}

\subsection{Fitting Details}

We make use of the Levenberg-Marquardt technique to solve the least-squares problem and to find the best solution for $m^{*}$, $\alpha$ and $\Phi^{*}$ using the parametrization given in equation \ref{eq:LF}. This method is known to be very robust and to converge even when poor initial parameters are given. However, it only finds local minima. In order to find the true global minimum and true best Schechter fit to our data we vary the starting parameters and choose the best fit solution. We show the Schechter fits for the for CARLA fields of richness $\geq 2\sigma$ in Fig. \ref{sample}. 

Our data are deep enough to not just fit for $m^{*}$ and $\Phi^{*}$ and assume a fixed $\alpha$ as had to be done in previous studies but to fit for all three quantities, $\alpha$, m$^{*}$ and $\Phi^{*}$ simultaneously. \citet{Mancone_2012} shows that $\alpha = -1$ fits well the galaxy cluster luminosity function at $1 < z < 1.5$ and that it stays relatively constant down to $z \sim 0$. We therefore also repeat the Schechter fits with fixed $\alpha = -1$ (Table \ref{tablecarla}).

\subsection{Uncertainty and Confidence Region Computation}

We estimate the uncertainties of the fitted parameters and the confidence regions of our fits using a Markov Chain Monte Carlo (MCMC) simulation. Fig. \ref{mcmc} shows the $1\ \sigma$ and $2\ \sigma$ contours for $\alpha$ and $m^*$ and the best Schechter fit.

We test our MCMC simulation by choosing a random set of $\Phi^*$, $\alpha$ and $m^*$ and computing $\Phi_i$ at $m_i$ following equation (\ref{eq:LF}). We then add normally distributed uncertainties $\Delta\Phi_{i}$ so that $\Phi_{i,err} = \Phi_i + \Delta\Phi_{i}$ and fit a Schechter function to $\Phi_{i,err}$. The original Schechter function $\Phi(\Phi^*,\alpha, m^*)$, the disturbed data points $\Phi_{i,err}$ and the Schechter fit agree very well with the 1 and 2 $\sigma$ confidence regions derived from the MCMC simulation and prove that the MCMC simulation provides us with a proper description of the uncertainties. Figure \ref{sample} shows the results of the Schechter fits to the CARLA clusters with $\Sigma_{\rm{CARLA}} > 2 \sigma$ for all redshift bins and the confidence regions derived from MCMC simulations.

\begin{table}
\tiny{
\caption{Schechter fit results for both $\alpha$ free and $\alpha$ fixed to $-1$ to the 4.5$\mu$m luminosity function. Below the horizontal line we also show Schechter fit results for the same analysis but taking SpUDS cells $\Sigma_{\rm{SpUDS}} = \Sigma_{\rm{SpUDS}}\pm1.6 \times \sigma_{\rm{SpUDS}}$ to estimate the background.}
\label{tablecarla}
\begin{center}
\begin{tabular}{l c |c c| c |c c}
\hline\hline
$\Sigma_{\rm{CARLA}}$ & $\langle{z}\rangle$ & $m^*_{4.5\mu m}$ & $\alpha$ & $m^*_{4.5 \mu m, \alpha = -1}$ & $N$  \\
\hline

   $> 2 \ \sigma$ &1.45       & $19.84       ^{+ 0.30      }_{-      0.20    }$ & $-1.01       ^{+ 0.10      }_{- 0.10     } $&$19.85       ^{+ 0.15      }_{- 0.15       }$&28 \\
     $> 2 \ \sigma$ &1.77       & $20.23       ^{+ 0.30      }_{-      0.20    }$ & $-1.07       ^{+ 0.15      }_{- 0.15     } $&$20.32       ^{+ 0.20      }_{- 0.15       }$&30\\
     $> 2 \ \sigma$ &2.05       & $20.19       ^{+ 0.30      }_{-      0.20    }$ & $-0.92       ^{+ 0.15      }_{- 0.15     } $&$20.08       ^{+ 0.20      }_{- 0.15       }$&30\\
     $> 2 \ \sigma$ &2.26       & $19.91       ^{+ 0.25      }_{-      0.25    }$ & $-1.28       ^{+ 0.15      }_{- 0.15     } $&$20.41       ^{+ 0.20      }_{- 0.15       }$&30\\
   $> 2 \ \sigma$ &2.51       & $19.96       ^{+ 0.35      }_{-      0.20    }$ & $-1.10       ^{+ 0.15      }_{- 0.15     } $&$20.10       ^{+ 0.20      }_{- 0.15       }$&30\\
      \vspace*{0.5cm}
     $> 2 \ \sigma$ &2.92       & $19.60       ^{+ 0.30      }_{-      0.20    }$ & $-1.29       ^{+ 0.10      }_{- 0.15     } $&$20.10       ^{+ 0.20      }_{- 0.15       }$&30\\
      
     $2.5 < \sigma < 3.5$ & 1.65       & $20.38       ^{+ 0.20      }_{-      0.20    }$ & $-0.75       ^{+ 0.20      }_{- 0.10     } $&$20.10       ^{+ 0.15      }_{- 0.15       }$&25\\
   $2.5 < \sigma < 3.5$& 2.23      & $19.99       ^{+ 0.20      }_{-      0.20    }$ & $-1.13       ^{+ 0.20      }_{- 0.15     } $&$20.21       ^{+ 0.15      }_{- 0.15       }$&27\\
           \vspace*{0.5cm}
$2.5 < \sigma < 3.5$ & 2.81       & $19.74       ^{+ 0.20      }_{-      0.20    }$ & $-1.17       ^{+ 0.15      }_{- 0.15     } $&$19.99       ^{+ 0.15      }_{- 0.15       }$&27\\
        
      $> $\ 3.5 $\sigma$ &1.49       & $19.88       ^{+ 0.20      }_{-      0.25    }$ & $-0.89       ^{+ 0.10      }_{- 0.10     } $&$19.72       ^{+ 0.15      }_{- 0.15       }$&18\\
      $> $\ 3.5 $\sigma$ &1.88       & $20.18       ^{+ 0.30      }_{-      0.20    }$ & $-1.01       ^{+ 0.20      }_{- 0.15     } $&$20.20       ^{+ 0.25      }_{- 0.15       }$&19\\
      \vspace{0.3cm}
      $> $\ 3.5 $\sigma$ &2.49       & $20.23       ^{+ 0.30      }_{-      0.20    }$ & $-0.95       ^{+ 0.20      }_{- 0.15     } $&$20.16       ^{+ 0.20      }_{- 0.15       }$&20\\
  \hline 
               $> $2 $\sigma$ &1.45       & $19.81       ^{+ 0.30      }_{-      0.20    }$ & $-1.02       ^{+ 0.10      }_{- 0.15     } $&$19.84       ^{+ 0.20      }_{- 0.15       }$&28\\
      $> $2. $\sigma$  &1.77       & $20.25       ^{+ 0.30      }_{-      0.20    }$ & $-1.07       ^{+ 0.15      }_{- 0.15     } $&$20.34       ^{+ 0.15      }_{- 0.15       }$&30\\
      $> $2 $\sigma$ & 2.05       & $20.16       ^{+ 0.30      }_{-      0.20    }$ & $-0.93       ^{+ 0.15      }_{- 0.15     } $&$20.07       ^{+ 0.20      }_{- 0.15       }$&30\\
      $> $2 $\sigma$  &2.26       & $19.86       ^{+ 0.25      }_{-      0.25    }$ & $-1.32       ^{+ 0.15      }_{- 0.10     } $&$20.44       ^{+ 0.20      }_{- 0.15       }$&30\\
      $> $2 $\sigma$  &2.51       & $19.91       ^{+ 0.20      }_{-      0.30    }$ & $-1.12       ^{+ 0.10      }_{- 0.15     } $&$20.09       ^{+ 0.20      }_{- 0.15       }$&30\\
                 \vspace*{0.5cm}
      $> $2 $\sigma$  &2.92       & $19.49       ^{+ 0.30      }_{-      0.25    }$ & $-1.33       ^{+ 0.15      }_{- 0.10     } $&$20.09       ^{+ 0.20      }_{- 0.15       }$&30\\

     $2.5 < \sigma < 3.5$ &1.65       & $20.39       ^{+ 0.20      }_{-      0.20    }$ & $-0.73       ^{+ 0.15      }_{- 0.15     } $&$20.10       ^{+ 0.15      }_{- 0.15       }$&25\\
       $2.5 < \sigma < 3.5$  &2.23       & $19.94       ^{+ 0.20      }_{-      0.20    }$ & $-1.16       ^{+ 0.15      }_{- 0.15     } $&$20.22       ^{+ 0.15      }_{- 0.10       }$&27\\
                 \vspace*{0.5cm}
     $2.5 < \sigma < 3.5$& 2.81       & $19.69       ^{+ 0.25      }_{-      0.15    }$ & $-1.19       ^{+ 0.15      }_{- 0.10     } $&$19.97       ^{+ 0.15      }_{- 0.15       }$&27\\
           
          $> $\ 3.5 $\sigma$  &1.49       & $19.86       ^{+ 0.25      }_{-      0.20    }$ & $-0.89       ^{+ 0.10      }_{- 0.10     } $&$19.71       ^{+ 0.15      }_{- 0.15       }$&18\\
      $> $\ 3.5 $\sigma$   &1.88       & $20.19       ^{+ 0.20      }_{-      0.20    }$ & $-1.01       ^{+ 0.10      }_{- 0.15     } $&$20.21       ^{+ 0.15      }_{- 0.15       }$&19\\
                 \vspace*{0.5cm}
       $> $\ 3.5 $\sigma$  &2.49       & $20.21       ^{+ 0.20      }_{-      0.20    }$ & $-0.95       ^{+ 0.15      }_{- 0.10     } $&$20.15       ^{+ 0.15      }_{- 0.15       }$&20\\
 
\end{tabular}
\end{center}}
\end{table}

\section{Robustness Tests}

\subsection{Stability of the IRAC Color Criterion with Redshift}

\begin{figure}
\centering
\includegraphics[scale = 0.4]{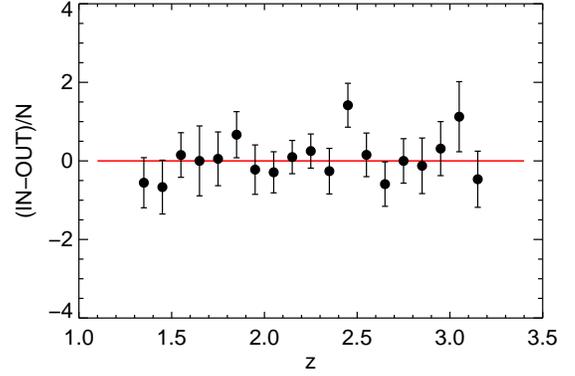}  
\caption{Average number of sources that could scatter in or out of the IRAC color selection due to photometric uncertainties as a function of candidate cluster redshift. `IN' are defined as sources currently not selected (e.g., $[3.6]-[4.5] < -0.1$), but which could easily scatter in due to photometric uncertainties, $[3.6]-[4.5] + \sigma{[3.6]-[4.5]} > -0.1$.  `OUT' are defined as currently selected sources (e.g., $[3.6]-[4.5] > -0.1$), but which could easily scatter out due to photometric uncertainties, $[3.6]-[4.5] - \sigma{[3.6]-[4.5]} < -0.1$. The results are relatively flat and centered at zero, implying that the color criterion used is efficient and stable at all redshifts probed here.}
\label{scattering}
\end{figure}

The evolution of the [3.6]-[4.5] color is very steep at $1.3 < z < 1.7$; cluster members in our lowest redshift bin are expected to have bluer colors, i.e. closer to $-0.1$, than cluster members at higher redshift. Faint sources which have a larger color uncertainty would therefore be expected to pass the criterion in some cases and not pass the criterion in other cases. For the contaminating background/foreground sources, this scattering is expected to be the same at all cluster redshifts. We therefore test if this effect has a statistically significant influence on the faint end of the luminosity function in the lowest redshift bin. Fig. \ref{scattering} shows the normalised number difference between sources where [3.6]-[4.5] $> -0.1$ but [3.6]-[4.5]-$\sigma_{[3.6]-[4.5]} < -0.1$ and sources where [3.6]-[4.5] $< -0.1$ but [3.6]-[4.5]+$\sigma_{[3.6]-[4.5]} > -0.1$. In the following we refer to these sources as `scatter-in sources' and `scatter-out sources', respectively. As the sum of the background and foreground color distribution is not expected to be dependent on cluster redshift and will therefore be the same for all redshift bins, any evolution in the difference between `scatter-in' and `scatter-out' sources will be caused by the cluster members. As Fig. \ref{scattering} shows, the evolution of the difference is consistent with being flat with redshift. The mean of the distribution is $0.06$ (N$_{\rm{fields}})^{-1}$ and a Spearman rank correlation analysis only gives a $33\%$ chance that there is an evolution of the difference with redshift. This test confirms that the color selection is very efficient and stable with redshift and that statistically the same portion of cluster members is selected at all redshifts.

\subsection{Validation of Luminosity Function Measurement Method Using ISCS Clusters}

We use the ISCS \citep{Eisenhardt_2008} to validate the methodology used in this paper and to compare to previously obtained results. \citet{Mancone_2010} uses the ISCS to derive luminosity functions, though he restricted the analysis to candidate cluster members based on photometric redshifts, computed as in \citet{Brodwin_2006} and with deeper photometry from the \textit{Spitzer} Deep, Wide-Field Survey \citep[SDWFS; ][]{Ashby_2009}. We measure the luminosity function around ISCS clusters at $1.3 < z < 1.6$ and $1.6 < z < 2$, i.e. overlapping in redshift with the CARLA cluster sample. We again first apply the IRAC color criterion to the sources in a cell with radius $r = 1$ arcmin around the cluster center to determine cluster member candidates and carry out a background subtraction using SpUDS as described above. We then fit a Schechter function to the resulting data with $\alpha$ fixed to $-1$. The SDWFS data is shallower than CARLA and we can therefore only measure the luminosity function down to [4.5]$=21.4$. The fitted values for $m^*$ at $4.5 \ \mu$m are $20.25^{+0.30}_{-0.30}$ and  $20.57^{+0.35}_{-0.35}$ for clusters at $1.3 < z < 1.6$ and $1.6 < z < 2$, respectively. The $m^*$ for the same redshift bins and $\alpha = -1$ found by \citet{Mancone_2010} are  $20.48^{+0.12}_{-0.09}$ and  $20.71^{+0.18}_{-0.12}$. Our results agree very well with these values, though note the larger error bars inherent to our simple color selection compared to the photometric redshift selection of \citet{Mancone_2010} which minimizes background contamination by incorporating extensive multi-wavelength supporting data. This test confirms that the method used in this paper, i.e. color-selecting galaxy cluster members and carrying out a statistical background subtraction, provides robust results. With this test we also confirm the trend towards fainter magnitudes for $m^{*}$ in ISCS clusters in the highest redshift bins reported by \citet{Mancone_2010}.

\begin{figure*}
\centering
\includegraphics[scale = 0.7]{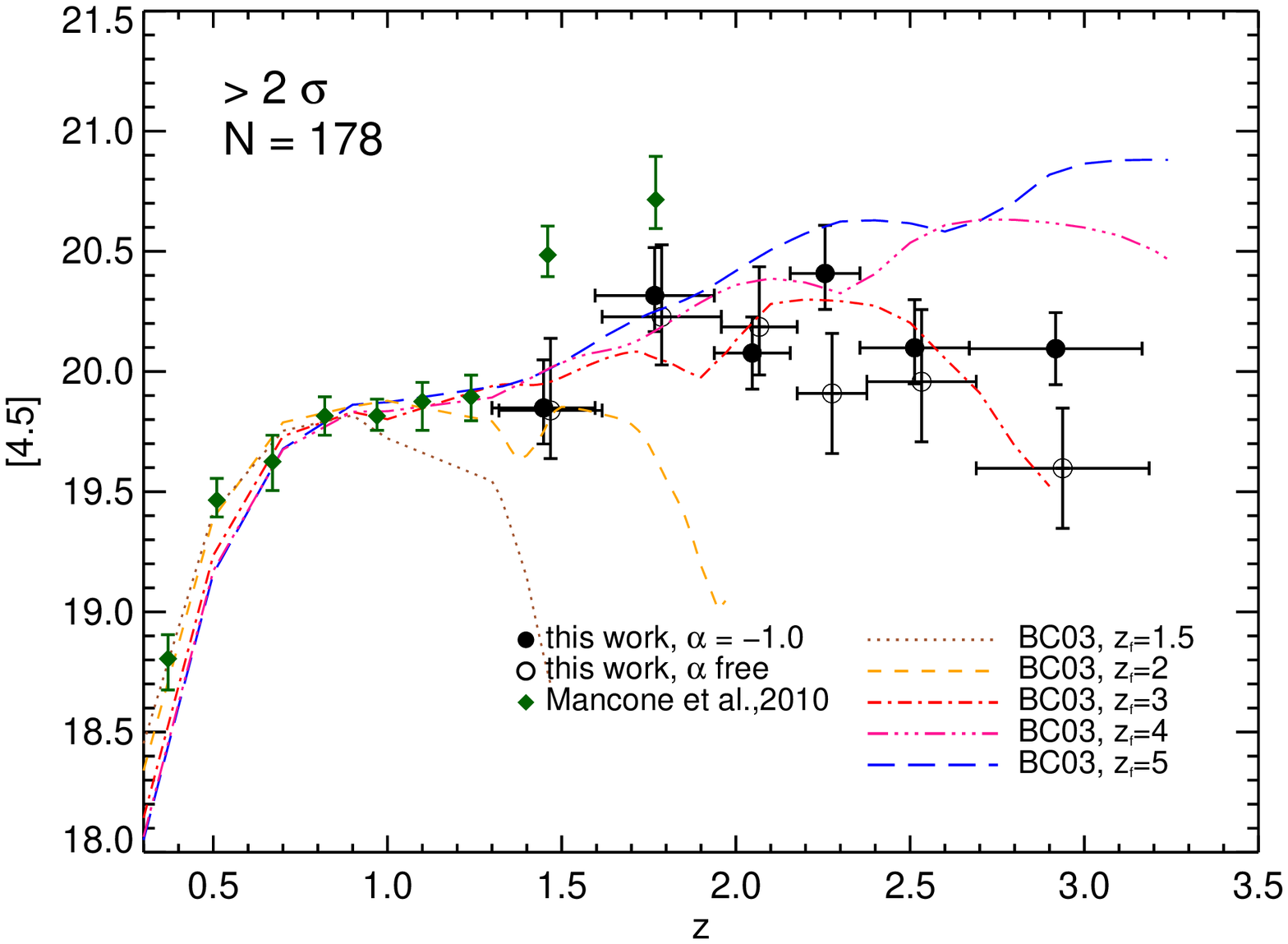}

\includegraphics[scale = 0.44]{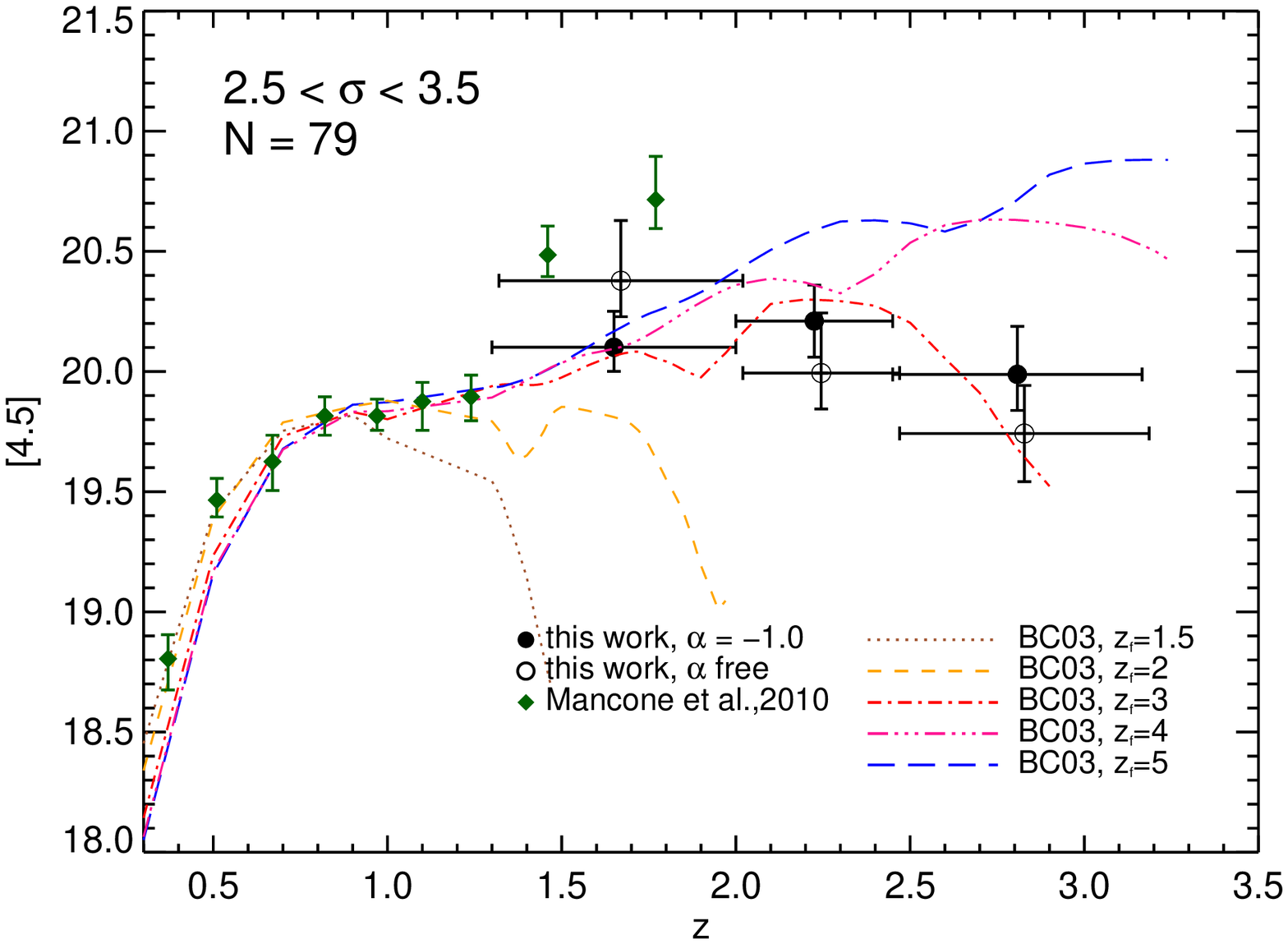}
\includegraphics[scale = 0.44]{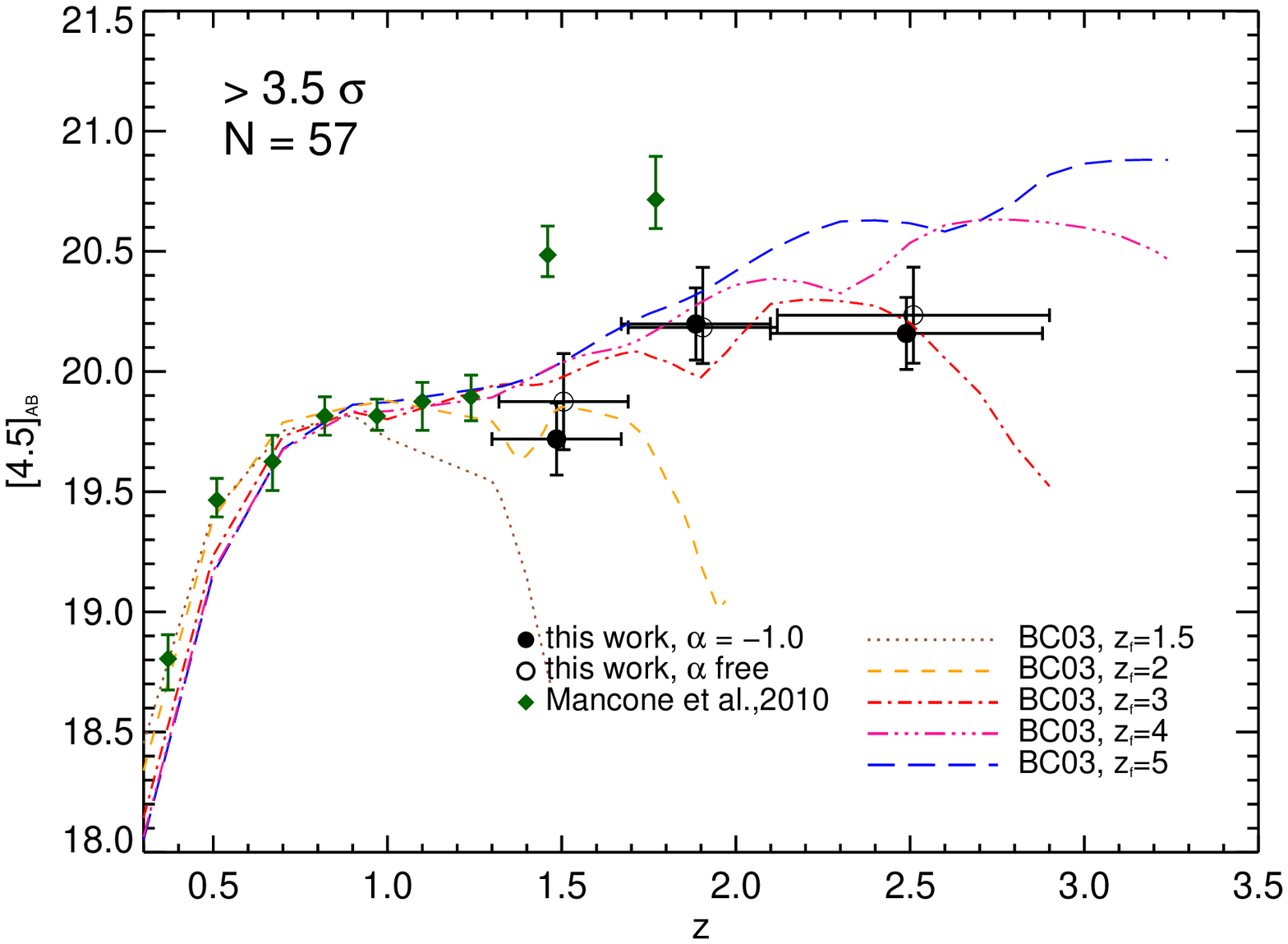}
\caption{Evolution of $m^*_{4.5 \mu m}$ with redshift for different CARLA cluster densities compared with results from \citet{Mancone_2010} and model predictions for passive stellar population evolution \citep{Bruzual_2003} with different formation redshifts, $z_f$. The number of CARLA fields going into each analysis is indicated on the plots. In most cases the results for the Schechter fits with fixed $\alpha$ and free $\alpha$ agree well. In cases where they don't this is due to bigger uncertainties at the faint end of the luminosity function in the data. In these cases the fixed $\alpha$ fit is probably the more meaningful result as it has been shown that generally a faint end slope of $\alpha = -1$ describes the data well at all CARLA densities and all redshifts. In general, our results are consistent with the passive evolution models out to $z \sim 3$ with formation redshifts $3 < z_f < 4$. The lower-density CARLA sample, shown in the lower left panel, seems to be consistent with results from \citet{Mancone_2012} at $z \leqslant 1.3$ and suggests that this CARLA subsample is similar to the clusters studied there. The high-density subsample of CARLA clusters, shown in the lower right panel are $\sim$ 0.7 mags fainter in the highest redshift bin compared to the lower richness samples. This might imply that the high richness sample consists of older clusters and therefore could provide a valuable sample of high-richness high-redshift clusters to study the formation of the earliest massive galaxies. For clarity, the results for a fixed $\alpha$ are slightly shifted along the x-axis.}
\label{m_evol}
\end{figure*}

\section{The Redshift Evolution of m$^{*}$}

\subsection {Comparison to Galaxy Evolution Models}

Figure \ref{m_evol} shows the evolution of $m^*$ with redshift in the context of passively evolving stellar population models. We also include results from \citet{Mancone_2010}. The measured $m^*$ values are compared to previous work and to model predictions of passively evolving stellar populations from \citet{Bruzual_2003} normalized to match the observed $m^*$ of galaxy clusters at $z \sim 0.82$, [4.5]$^{*}= 19.82^{+0.08}_{-0.08}$ (AB) for $\alpha = -1$ \citep{Mancone_2012}. Note that the results and implications of our work are independent of the model normalisation. Although \citet{Mancone_2010} used a normalisation obtained from lower redshift cluster analyses we use a normalisation at relatively high redshift to match the redshift of the CARLA data as best as possible. 

\citet{Mancone_2010} carried out a similar study and measured the IRAC1 and IRAC2 luminosity function for 296 galaxy clusters from SDWFS over $0.3 < z < 2$. The clusters were identified as peaks in wavelet maps generated in narrow bins of photometric redshift probabilities. At $z < 1.3$, the evolution of $m^*$ agreed well with predictions from passively evolving stellar populations and no mass assembly, with an inferred formation redshift of $z_f = 2.4$. For the two highest redshift bins at $z > 1.3$, the results disagreed significantly from the continuation of the passive evolution model. \citet{Mancone_2010} interpreted this deviation as possible ongoing mass assembly at these epochs. The results here for the CARLA clusters do not agree with nor continue the trend \citet{Mancone_2010} report at $z > 1.3$. In the redshift bins in which our study and \citet{Mancone_2010} overlap, we find $m^*$ to be $\sim 0.5$ mag brighter and to continue the trend found at lower redshifts by \citet{Mancone_2010}. At lower redshift ($1.3 < z < 1.8$) the models do not differ much and $m^{*}$ is therefore consistent with a range of formation redshifts. At higher redshift the models diverge and show that clusters may have formed early with formation redshifts in the range $3 < z_{f} < 4$. 

At all redshifts our results are consistent with passive evolution models with $3 < z_{f} < 4$ . As CARLA clusters at $z \sim 3$ will not necessarily be progenitors of CARLA clusters at $z \sim 1.5$, it does not necessarily imply that they will remain passive subsequently. However, even at the highest redshifts probed here ($z \sim 3$) we do not measure a prominent deviation from passive evolution models nor do we see signs of significant mass assembly.

\subsection{Dependence on Cluster Richness}

We also study the evolution of $m^*$ for two CARLA subsamples of different richnesses ($2.5 \ \sigma < \Sigma_{\rm{CARLA}} < 3.5 \ \sigma$ and $\Sigma_{\rm{CARLA}} > 3.5 \ \sigma$) and plot the results in Fig. \ref{m_evol}. Although it is not fully known how the richness of IRAC-selected sources scales with mass, we assume here that statistically these two subsamples will represent clusters of increasing mass. For the lower richness sample, the evolution of $m^{*}$ appears to be similar to the full sample with formation redshifts of  $3 < z_{f} < 4$. Excluding the highest and lowest richness clusters therefore does not seem to have a significant effect on the luminosity functions. At lower redshift (i.e. $1.3 < z < 2.0$) the result for $m^{*}$ and free $\alpha$ is consistent with results from \citet{Mancone_2010} at $z \leqslant 1.3$ which might suggest that the CARLA lower richness clusters at this redshift are similar to the ISCS clusters studied \citet{Mancone_2010}.

Only considering the highest richness sample of CARLA clusters results in a monotonically increasing $m^{*}$, with $m^{*}$ $\sim$ 0.7 mags fainter in the highest redshift bin compared to the lower richness samples. This might imply that the high richness sample includes slightly older stellar populations. This subsample is thus of particular interest for future follow-up as this population of rich, high-redshift clusters could provide a powerful probe to study the early formation of massive galaxies in the richest environments and --- from a cosmological point of view --- test the predictions of $\Lambda$CDM \citep{Brodwin_2010}.

\subsection{Difference between $\alpha = free$ and $\alpha = -1$ fits }

As can be seen in Fig. \ref{mcmc}  and Fig. \ref{m_evol}, the Schechter fit results for $\alpha$ as a free parameter and $\alpha$ fixed to $-1$ generally agree well within the $1 \sigma$ uncertainties. For the few exceptions, the fits do agree within the $2 \sigma$ uncertainties. We conclude that a fixed $\alpha$ of $-1$ describes the luminosity function functions well at all redshifts and all densities probed in this paper. With deeper IRAC observations (1400 s of exposure time), albeit not as deep as the CARLA observations (2000 s of exposure time) and deep multi-wavelength supporting data, \citet{Mancone_2012} study a subset of the original B\"{o}otes sample and measure the faint end slope $\alpha$ of the mid-IR galaxy cluster luminosity function to be $\sim -1$ at $1 < z < 1.5$. Similar studies at lower redshift measure similar slopes. This suggests that the shape of the cluster luminosity function is mainly in place at $z = 1.3$.

The luminosity functions measured in this paper are consistent with $\alpha = -1$ at all redshifts and all richnesses probed. Combined with results from \citet{Mancone_2010} and \citet{Mancone_2012}, this result suggests that galaxy clusters studied in this paper have already started to assemble low-mass galaxies at early epochs. Further processes that govern the build up of the cluster and that are discussed in more detail in \S~6 then have probably no net effect on the shape of the luminosity function. This is consistent with the results found by Mancone et al. (2010).

\section{Discussion}

\subsection{Alternatives to pure passive Evolution Models}

The above results suggest an early formation epoch for galaxy clusters that are passively evolving and an early build-up of the low-mass galaxy population. These measurements, however, seem to be at odds with results investigating lower redshift clusters. \citet{Thomas_2010} shows that the age distribution in high-density environments is bimodal with a strong peak at old ages and a secondary peak comprised of young, $\sim 2.5$ Gyr old galaxies. This secondary peak contains about $\sim 10$\% of the objects. Similarly, \citet{Nelan_2005} derives a mean age of low-mass objects of low redshift galaxy clusters to be about 4 Gyrs in low-redshift clusters. Although their observation suggests a decline of star-forming galaxies and a trend of downsized galaxy formation, low mass galaxies are still assembling until relatively recent times. Measurements of the star-formation activity for higher redshift cluster galaxies also provide evidence for continuous star-formation activity, albeit evolving more rapidly than the star-formation activity in field galaxies \citep[e.g.]{Alberts_2014}.

We therefore investigate the extent to which our results can be explained by a sum of various stellar populations to estimate the maximum fraction of a star-forming cluster population that is still consistent with the data. We divide the cluster population into two stellar populations, a simple stellar population (SSP) with a delta-burst of star formation at high redshift and a composite stellar population (CSP) with a continuous, only slowly decaying star-formation rate of the form $\propto exp(-t/\tau)$ and large $\tau$. The details of the three different model sets we derive are as follows:

\begin{itemize}
\item Model 1: Sum of SSP with $z_f = 3$ and CSP with $z_f = 3$ and $\tau =  10$ Gyr with ratios (SSP:CSP) ranging between 90:10 and 0:100\footnote{a ratio (SSP:CSP) of (100:0) is equivalent with a passive evolution model and is already discussed above} by mass.
\item Model 2: Sum of SSP with $z_f = 3$ and CSP with $z_f = 3$ and $\tau =  1$ Gyr with ratios (SSP:CSP) ranging between 90:10 and 0:100 by mass.
\item Model 3: Sum of SSP with $z_f = 5$ and CSP with $z_f = 3$ and $\tau =  10$ Gyr with ratios (SSP:CSP) ranging between 90:10 and 0:100 by mass.
\end{itemize}

A prolonged mass assembly means that at high redshift the observed $4.5\ \mu$m magnitude of galaxies is fainter because the stellar population is still forming. Therefore, accounting for mass assembly, i.e. allowing for star-forming population to contribute to the observed $m^{*}$, causes $m^{*}$ to become fainter at high redshift depending on the contribution of this star-forming population.

\begin{figure}
\centering
\includegraphics[scale = 0.43]{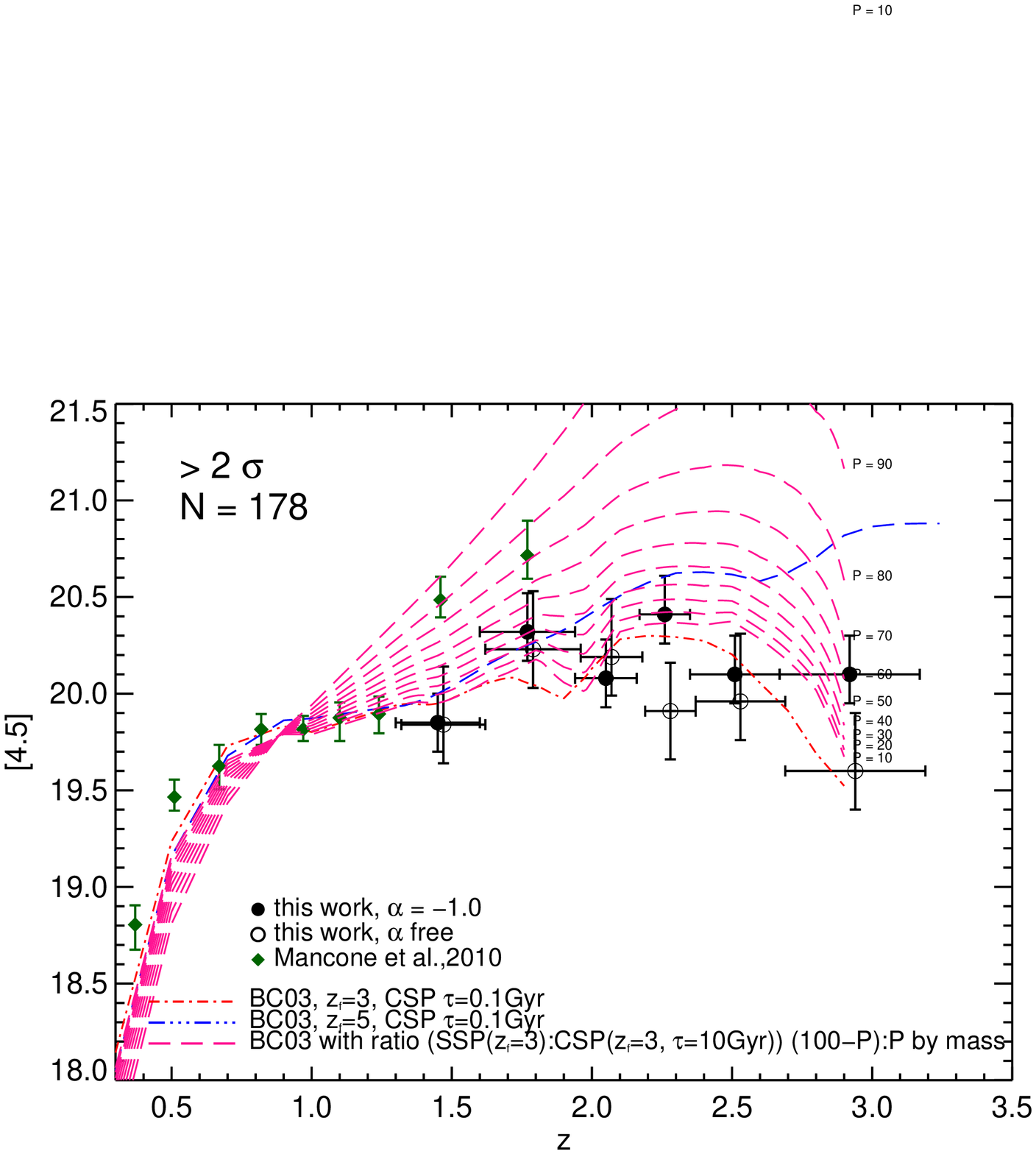}  
\includegraphics[trim = 0 0 0 2 cm, clip = true, scale = 0.43]{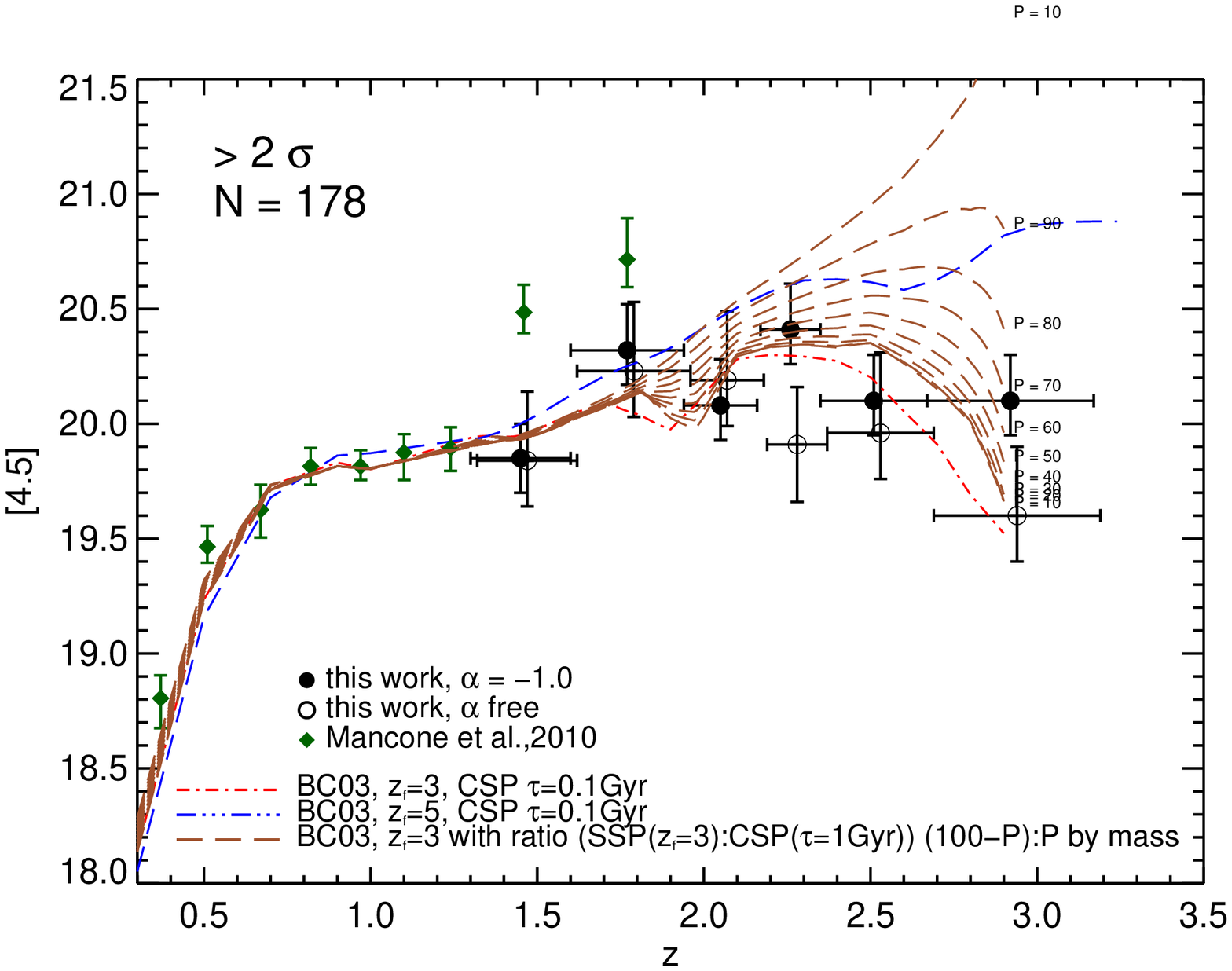}
\includegraphics[scale = 0.43]{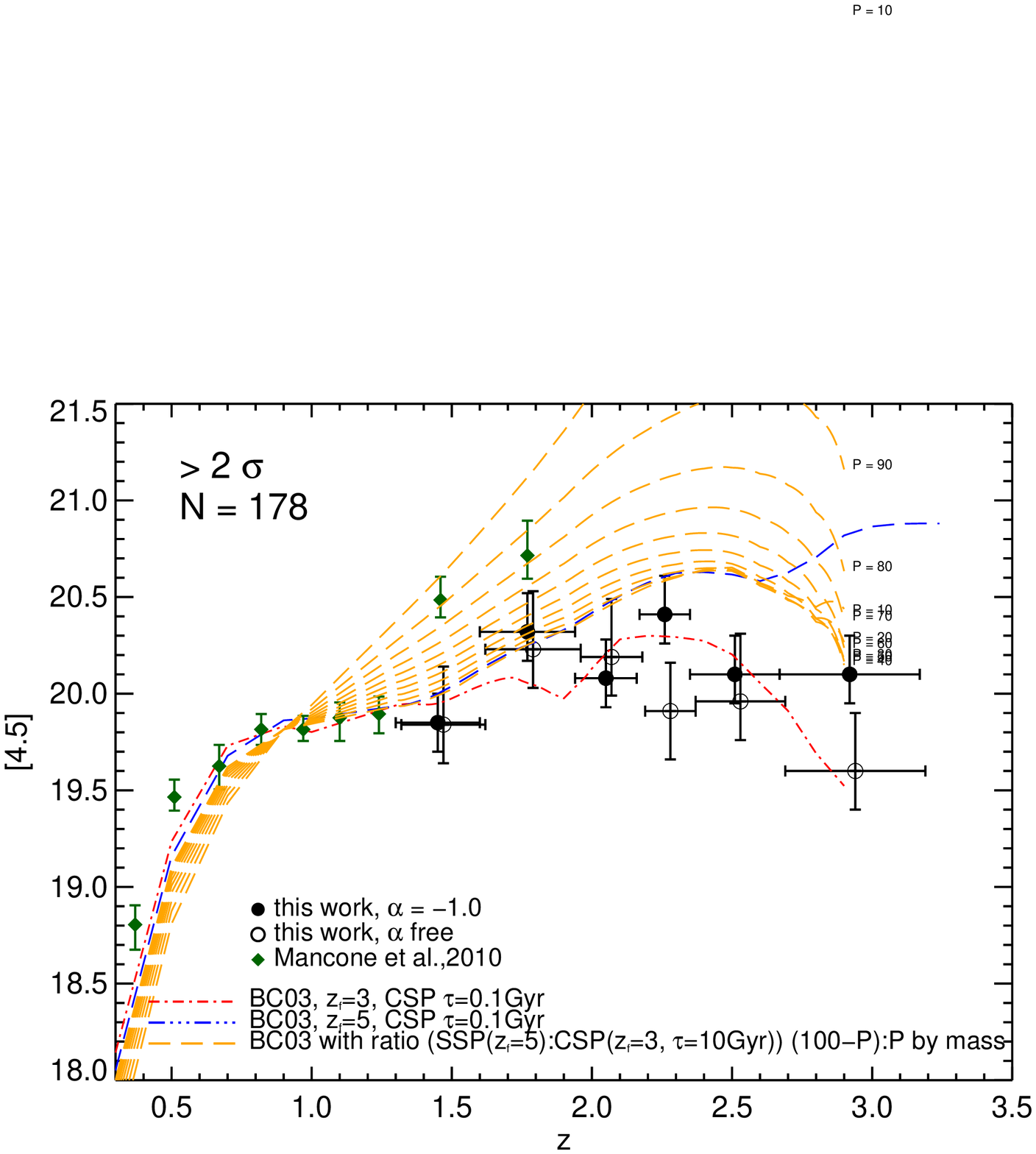}
\caption{Model predictions for the evolution of $m^*$ for a superposition of a passive and star-forming galaxy population. The passive, simple stellar population (SSP) formed stars in a delta burst at $z_f$ and evolved passively thereafter while the star-forming, composite stellar population (CSP) shows an exponentially decaying SFR with an $e$-folding timescale $\tau$. We show models with differing mass ratios of the SSP and CSP. Comparing the models with our measurements for $m^{*}$ allows to set an upper limit on the contribution of the CSP. \textit{Top panel: Model 1:} SSP with $z_f = 3$, CSP with $z_f = 3$ and $\tau = 10$ Gyrs \textit{Middle panel: Model 2:} SSP with $z_f = 3$, CSP with $z_f = 3$ and $\tau = 1$ Gyr  \textit{Bottom panel: Model 3:} SSP with $z_f = 5$, CSP with $z_f = 3$ and $\tau = 10$ Gyrs }
\label{models}
\end{figure}

In Figure \ref{models} we show these models in the context of our results. They allow us to set an upper limit on the star-forming fraction, $P$, in our candidate cluster sample. Model 1, which allows for a significant star formation that is only very slowly decaying with cosmic time, shows that at all redshifts the mass fraction of the star-forming population cannot be larger than 40\% (with one outlying exception of up to 60\% at $z \sim 1.7$). In Model 2 the star formation rate (SFR) of the CSP decays faster and the contribution of the SSP becomes dominant much earlier than in Model 1. It therefore predicts $m^{*}$ to be brighter at high redshifts and resemble the prediction of the passive evolution model. Consequently, our empirical results are in agreement with large contributions of up to 90\% of the CSP in Model 2, with the CSP starting to passively evolve $\sim 2.3$ Gyrs earlier (at $z \sim 1.7$) than in Model 1 (at $z \sim 0.9$). Model 3 shows the evolution of a mixed population with a delta burst of star formation at $z_{f} = 5$ (SSP) and a long burst of star formation at $z_f = 3$ (CSP with $\tau = 10$ Gyrs). The SSP with $z_f = 5$ is fainter in IRAC2 at $1.5 < z < 3$ than a SSP with $z_f = 3$, so that an additional starburst at $z=3$ leads to an even fainter $m^{*}$ at $1 < z < 3$ for Model 3. Model 3 does not reproduce the results of the LF analysis and therefore this scenario can be ruled out by the data.

This shows that the results for the evolution of $m^{*}$ obtained in this analysis -- although consistent with passive evolution models -- also allow for a limited contribution of a star-forming population in galaxy clusters. Our models show that this contribution is small (up to 40 \% by mass, but probably on average around $\sim 20$ \%) for a population with a high and slowly decaying star-formation rate, or that this contribution is large (up to 80\%) for a population with a fast decaying SFR and an evolution that resembles passive evolution $\sim 2.3$ Gyrs earlier.

For a sample of 10 rich clusters at $0.86 < z < 1.34$ \citet{vanderburg_2013} compares the contributions of quiescent and star-forming populations to the total mass function. We integrate the published mass functions for the quiescent and star-forming population \citep{vanderburg_2013} over galaxy masses with $10^{10.1} < M_*/M_{\sun} < 10^{11.5}$. We find a mass fraction of the quiescent population of $\sim 80$\% compared to the total stellar mass of the clusters . This is also in agreement with the upper limit for the fraction of the star-forming population derived for CARLA clusters in the lowest redshift bin. 

As CARLA clusters at $z \sim 3$ will not necessarily be the progenitors of CARLA clusters at $z \sim 1.5$, we unfortunately cannot constrain the evolution of the maximum fraction of a star-forming population and cannot make conclusions about the quenching timescales and processes. 

\subsection{Biases of the CARLA cluster sample}

Our analysis shows that the evolution of the CARLA clusters seems to be significantly different from the cluster sample analysed by \citet{Mancone_2010}. Using \textit{Spitzer} $24\ \mu$m imaging for the same cluster sample, \citet{Brodwin_2013} analysed the obscured star formation as a function of redshift, stellar mass and clustercentric radius. They find that the transition period between the era where cluster galaxies are significantly quenched relative to the field and the era where the SFR is similar to that of field galaxies occurs at $z \sim 1.4$. Combining these measurements with other independent results on that sample \citep{Snyder_2012, Martini_2013, Alberts_2014}, the authors conclude that major mergers contribute significantly to the observed star formation history and that merger-fueled AGN feedback may be responsible for the rapid truncation between $z = 1.5$ and $z = 1$. 

While the ISCS clusters studied in \citet{Mancone_2010} and \citet{Brodwin_2013} were selected from a field survey as 3-D overdensities using a photometric redshift wavelet analysis, the clusters studied here are found in the vicinity of RLAGN. With RLAGN belonging to the most massive galaxies in the universe \citep[$ m \sim 10^{11.5}$ M$_{\sun}$; ][]{Seymour_2007, Breuck_2010}, these clusters could reside in the largest dark matter halos, deepest potential wells and densest environments. 

Indeed, \citet{Mandelbaum_2009} derives halo masses for 5700 radio-loud AGN from the Data Release 4 of the Sloan Digital Sky Survey and finds the halo masses of these radio-loud AGN to be about twice as massive as those of control galaxies of the same stellar mass. Previous work \citep[e.g.][]{Best_2005}has shown that more massive black holes seem to trigger radio jets more easily, but as this boost in halo mass is independent of radio luminosity, the authors conclude that the larger-scale environment of the RLAGN must play a crucial role for the RLAGN phenomenon. Similarly, albeit at higher redshift, Hatch et al., in prep. finds that the environments of CARLA RLAGN are significantly denser than similarly massive quiescent galaxies. They detect a weak positive correlation between the black-hole mass and the environmental density on Mpc-scales, suggesting that even at high redshift the growth of the black hole is also linked to collapse of the surrounding cluster. 

This peculiar interplay between radio jet triggering, stellar mass, black hole and halo mass of the RLAGN and the larger-scale environment suggest that (proto-)clusters and the large-scale environments of RLAGN are distinct from clusters found in field surveys.

If mergers are significantly contributing to the transition of clusters from unquenched to quenched systems, as suggested in \citet{Brodwin_2013}, this transition redshift will be dependent on cluster halo mass. If the environments and dark matter halos around radio-loud AGN are indeed more massive this would explain the conflict of our results with those from \citet{Mancone_2010}. CARLA clusters have probably undergone this transition period much earlier than the ISCS clusters. At $1.4 < z <1.8$ where the star-forming fraction of ISCS clusters analysed by \citet{Mancone_2010} still seems to be very high ($\sim 80$\%), this contribution is already much smaller in CARLA clusters. 
As mentioned earlier, our measurements allow us to derive upper limits on the contribution of a star-forming population but do not allow for a definite constraint on the transition redshift.

\section{Summary}

We have shown the evolution of the mid-IR luminosity function for a large sample of galaxy cluster candidates at $1.3 < z < 3.2$ located around RLAGN. All cluster candidates in this work were identified as mid-IR color-selected galaxy overdensities, a selection that is independent of galaxy type and evolutionary stage. \citet{Wylezalek_2013} has shown that indeed these excess color-selected sources are centered on the RLAGN, implying they are associated. There is a steep increase of density toward the RLAGN which would not be seen if there was not a physical link between the galaxies in the field and the AGN. We therefore expect most of these overdensities to be true galaxy (proto)-clusters. Having neither spectroscopic nor photometric redshifts at hand, this study relies on statistics with 10 to 30 clusters per redshift bin. 

We have shown that our results are consistent with theoretical passive galaxy evolution models up to $z = 3.2$. \citet{Mancone_2010} measured significant deviation from these passive galaxy evolution models in their highest redshift bins ($1.3 < z < 1.8$), which correspond to our lowest redshift bins. They attributed this to possible ongoing mass assembly at these ($1.3 < z < 1.8$) redshifts. The work in this paper fails to confirm this previously observed trend but, on the contrary, finds that the cluster luminosity function agrees well with passive evolution models. To test our analysis, we apply our methodology to the clusters in \citet{Mancone_2010} and confirm their results at $1.3 < z < 1.8$. For lower richness CARLA clusters our results in the lowest redshift bins ($1.3 < z < 1.8$, free $\alpha$) are consistent with what has been found in \citet{Mancone_2010} for their $z < 1.3$ clusters and shows that the lower richness CARLA clusters are probably more similar to the clusters analysed in \citet{Mancone_2010}. 

We construct three different sets of galaxy evolution models to test to which extent our results are consistent with a star-forming population of galaxies contributing to $m^{*}$. This allows us to obtain upper limits on the mass fraction of a star-forming population. This modelling shows that at $z < 1.5$ the CARLA clusters are mostly passively evolving with very little contribution of a star-forming population. 

At $1.5 < z < 3$ the LF measurements are consistent with a star-forming population contributing to the cluster LF. This contribution is very likely low, i.e. less than 40\% by mass, for a star-forming population with large (10 Gyr) $e$-folding timescale. 

For star-forming populations with short (1 Gyr) $e$-folding timescale the modelling also allows for larger mass fractions. But due to the short $e-$folding timescales these models are effectively passively evolving.

Because CARLA clusters in the highest redshift bins ($z \sim 3$) are likely not progenitors of those at lower redshift ($z \sim 1.5$) and because uncertainties in $m^{*}$ are large we cannot constrain the evolution of the mass fraction of the star-forming population. This study shows, however, that the fraction of star-forming galaxies in CARLA clusters must be significantly smaller than in clusters studied in \citet{Mancone_2010}.

It has been shown that RLAGN seem to reside in dark matter halos twice as massive and denser galaxy environments as quiescent galaxies of similar stellar mass \citep{Mandelbaum_2009}. Recent work on the evolution of the star formation in ISCS clusters by \citet{Brodwin_2013} also suggests that the epoch of mass assembly should be considerably higher for high-redshift high-mass systems like SPT-CL J0205-5829 at $z = 1.32$ \citep{Stalder_2013} or XMMU J2235.3-2557 at $z = 1.39$ \citep{Mullis_2005}. These clusters have very low central star formation rates \citep[e.g. ][]{Stalder_2013} and already seem to be largely quenched and passive in their cores. In agreement with this prediction, CARLA clusters appear to have undergone this transition period to largely quiescent systems much earlier than ISCS clusters from \citet{Mancone_2010}. This can explain the discrepancies in the $m^*$ measurements for ISCS clusters and CARLA clusters at $z \sim 1.5$.

While the lower richness CARLA clusters are probably more similar to clusters found in \citet{Mancone_2010}, the high-richness CARLA clusters likely represent the high richness portion of the cluster population and therefore could provide a powerful cluster sample to study the formation of high-redshift, rich clusters and test predictions of the standard $\Lambda$CDM universe \citep[e.g., ][]{Mortonson_2011}.

Another key result of our work is the agreement of the data with of a relatively flat luminosity function ($\alpha = -1$) out to the highest redshifts. \citet{Mancone_2012} had already shown that at least up to a redshift of $z = 1.5$ the faint end slope of the luminosity function seems to be already in place and does not evolve since then. This implies that processes that are responsible for the build-up of the mass of low-mass cluster galaxies do not have any net effect on the the overall slope of the luminosity function. Processes that could steepen the slope of the luminosity function are star formation and mergers. Processes that could flatten the slope are galaxy-galaxy interactions, galaxy harassment or the dissolution of cluster galaxies \citep[see][and references therein for more extensive discussion]{Mancone_2012}. In this work, we find that a slope of $\alpha = -1$ describe the luminosity functions very well even to the highest redshifts probed in this work ($z = 3.2$). Processes that could steepen or flatten the faint end slope of the luminosity function seem not to have a significant net effect up to $z \sim 3$. Low mass galaxies seem to grow through star formation, get quenched and are replenished by in-falling field galaxies at a rate that does not have a major effect on the shape of the LF. 

This study provides a first statistical approach to measuring the high-redshift cluster LF, albeit biased to cluster candidates found in the fields of RLAGN. We have successfully started the follow-up of our most promising candidates and recently confirmed a structure around the radio galaxy MRC 0156-252 at $z = 2.02$ \citep{Galametz_2013}. We also started multi-object near-IR spectroscopic follow-up to confirm additional high-redshift clusters. Obtaining clean samples of spectroscopically confirmed galaxy clusters at high redshift with different techniques with different selection biases is necessary to fully probe the different scenarios and paths that galaxy clusters take in their evolution.

\acknowledgments

We thank the referee for helpful comments that have improved the manuscript. We gratefully thank Mark Lacy for allowing us to access SERVS images and catalogs and Roberto Assef and Conor Mancone for helpful discussions and advice. N. Seymour is the recipient of an ARC Future Fellowship. This work is based on observations made with the \textit{Spitzer Space Telescope}, which is operated by the Jet Propulsion Laboratory, California Institute of Technology under a contract with NASA.

\appendix

\begin{figure*}
\centering
\includegraphics[scale = 0.4]{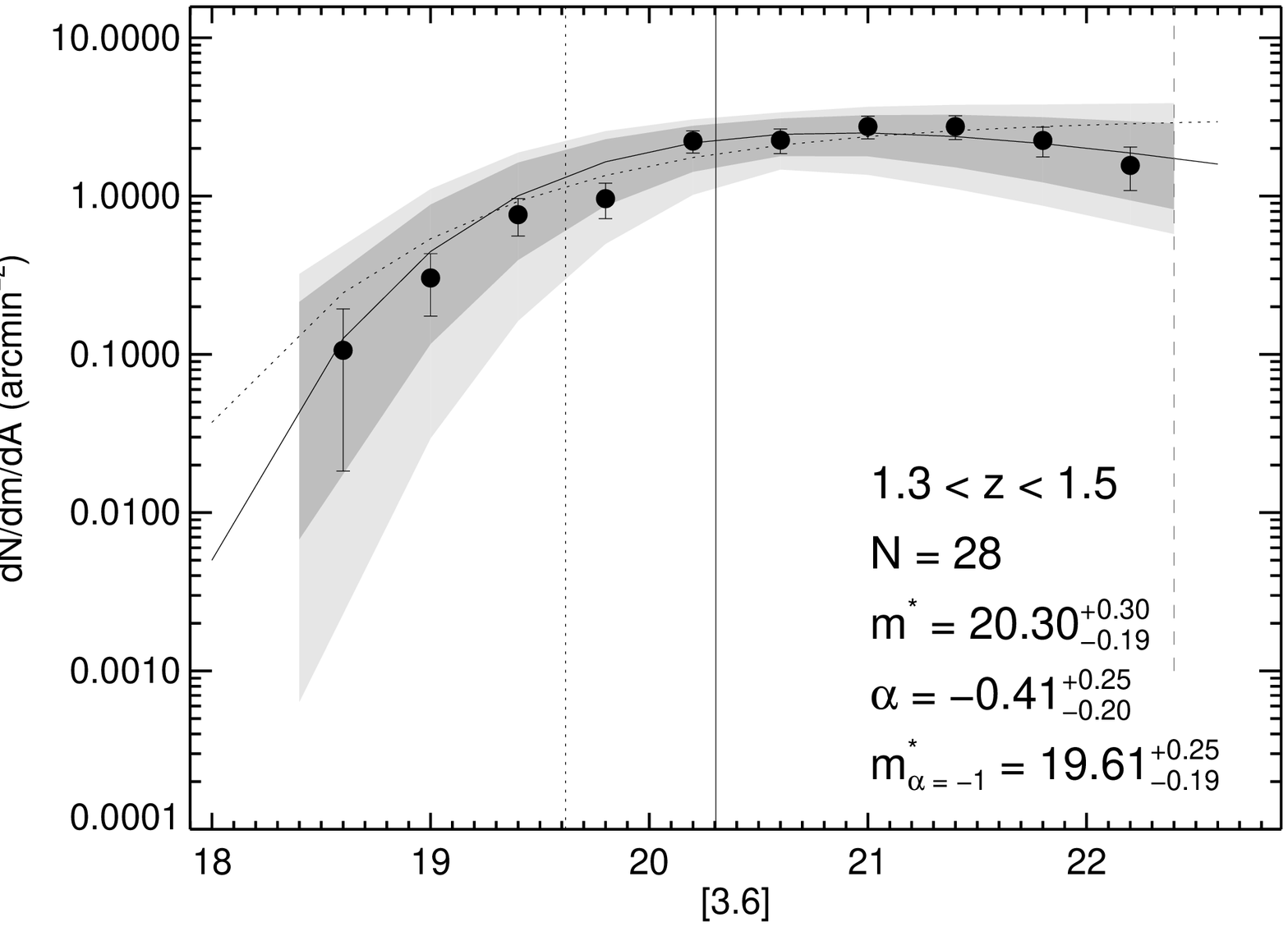}
\includegraphics[scale = 0.4]{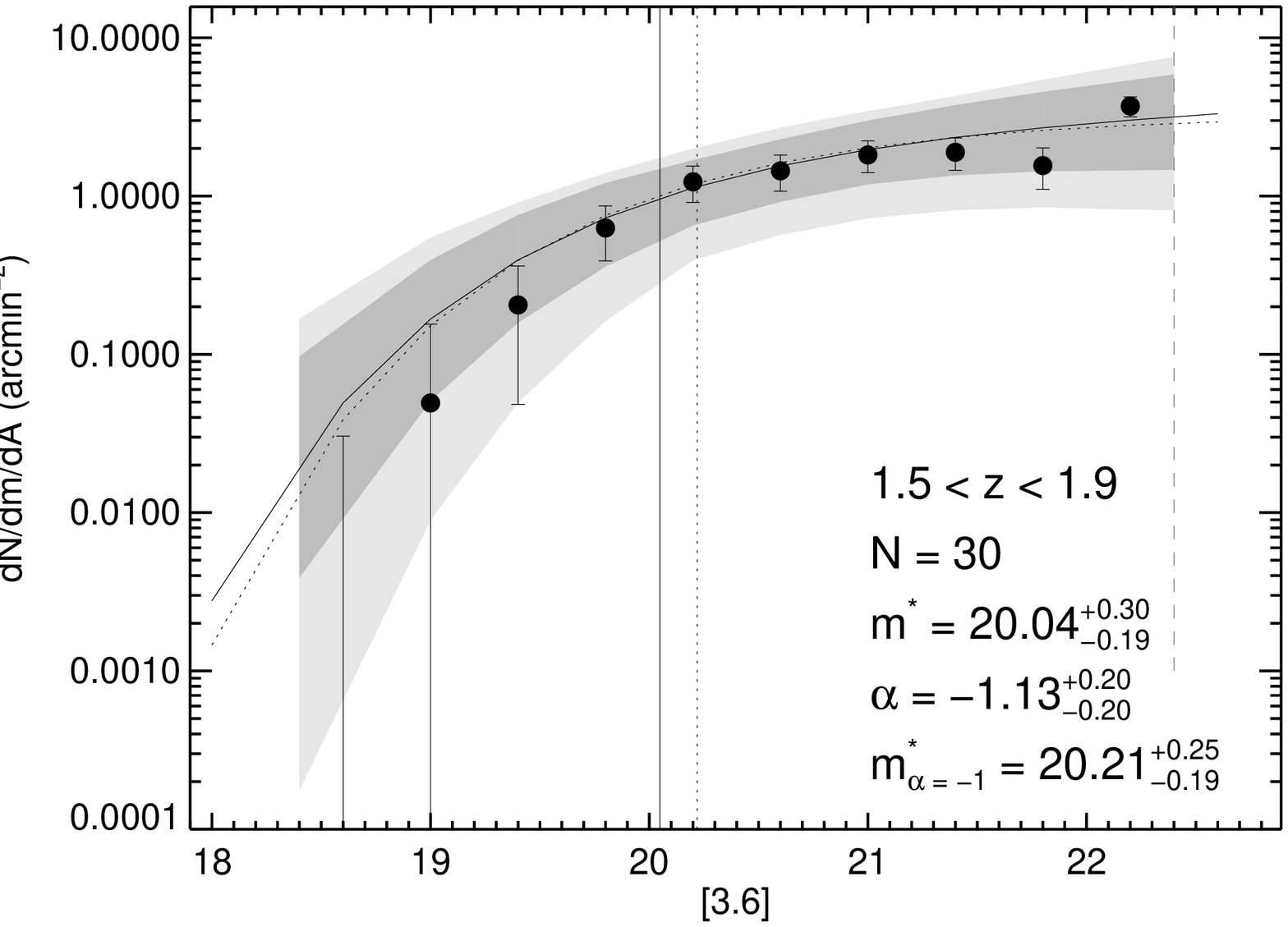}

\includegraphics[scale = 0.4]{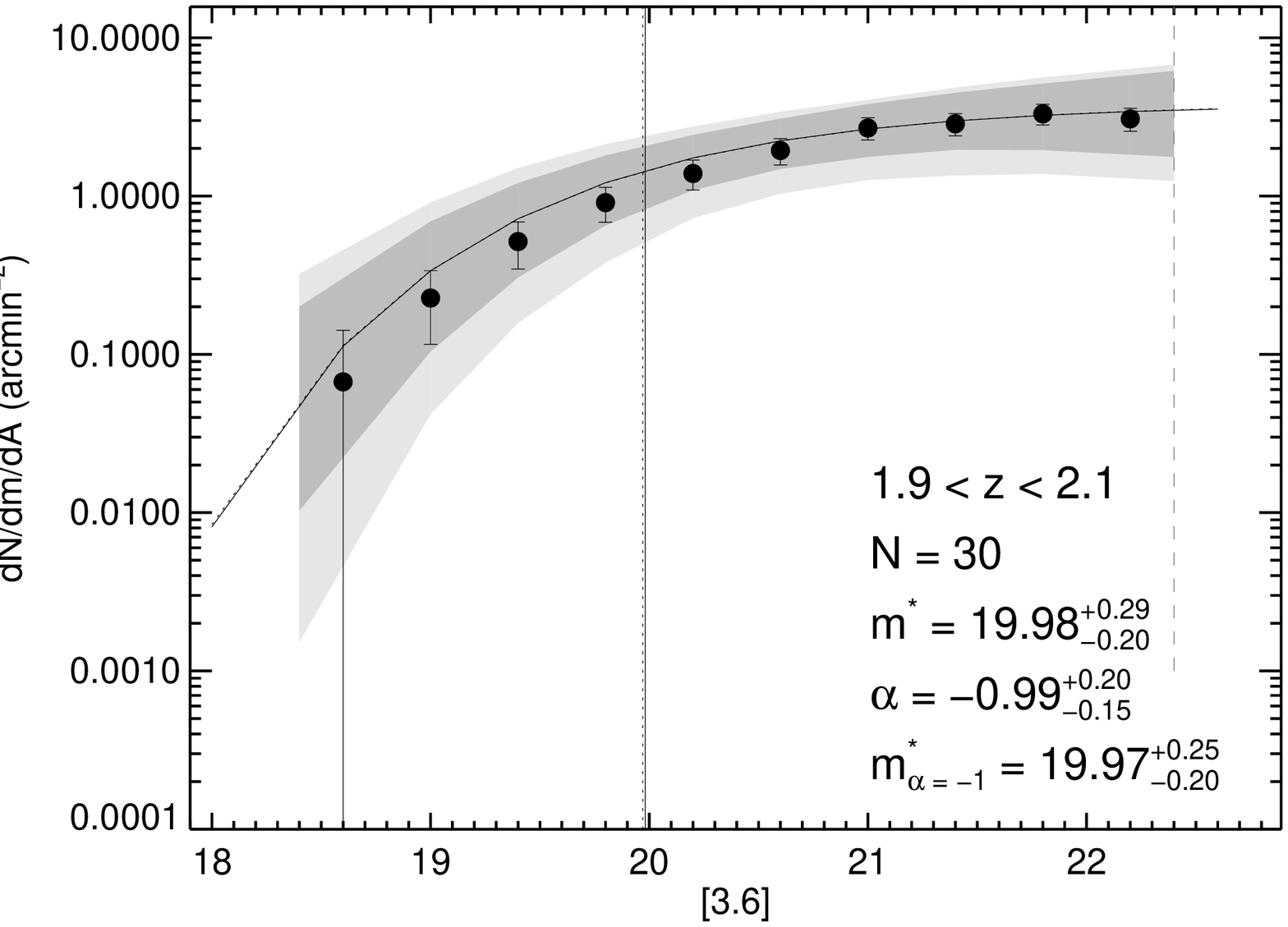}
\includegraphics[scale = 0.4]{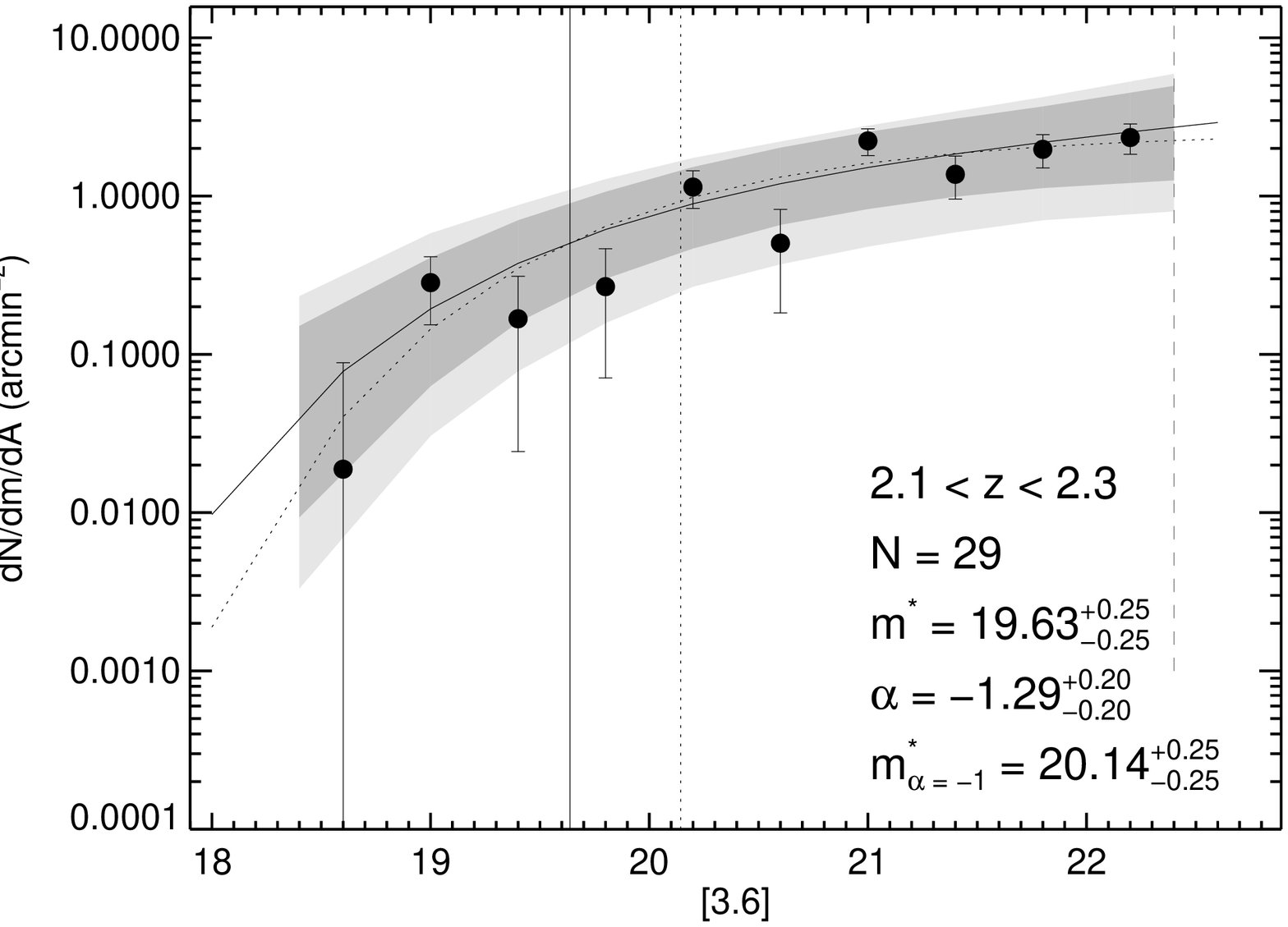}

\includegraphics[scale = 0.4]{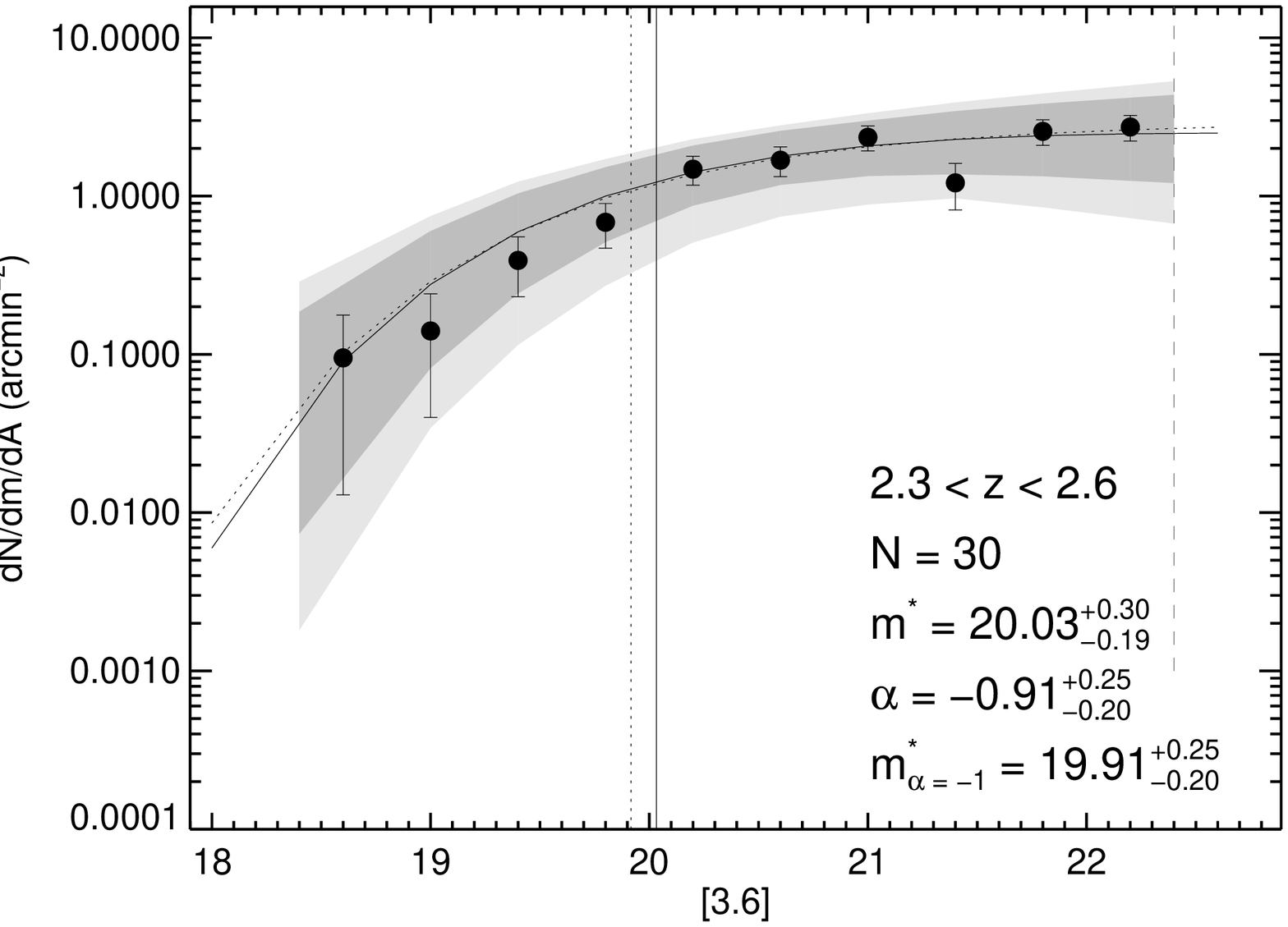}
\includegraphics[scale = 0.4]{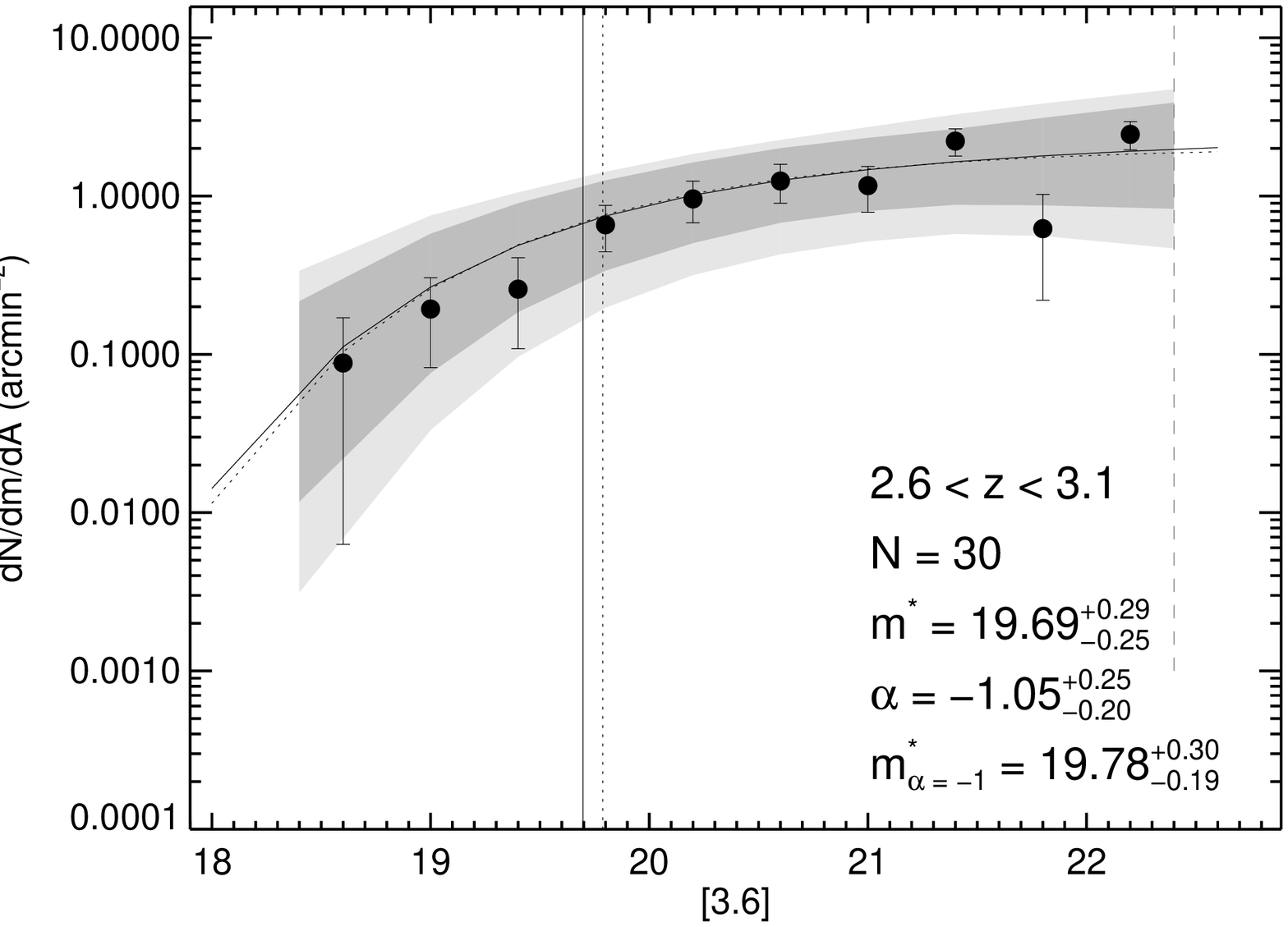}
\caption{Schechter fits to the 3.6 $\mu$m cluster luminosity function in each redshift bin for CARLA cluster members with $\Sigma_{\rm{CARLA}} > 2 \sigma$. The redshift bins were chosen to contain similar numbers of objects, $N$. The solid circles are the binned differences between the luminosity function for all IRAC-selected sources in the clusters and the background luminosity function derived from the SpUDS survey. The solid curve shows the fit to the data for a free $\alpha$ while the dotted curve shows the fit for $\alpha = -1$. The vertical solid and dotted lines show the fitted values for $m^{*}$ with free and fixed $\alpha$, respectively. The dark and light grey shaded regions show the $1 \sigma$ and $2 \sigma$ confidence regions for the $\alpha$-free-fit derived from Markov-Chain Monte Carlo simulations. The vertical dashed line shows the apparent magnitude limit. }
\label{sample_irac1}
\end{figure*}

Figure \ref{sample_irac1} shows the results of Schechter fits to cluster member candidates detected above the IRAC1 95\% completeness limit of 3.45 $\mu$Jy for CARLA clusters with $\Sigma_{\rm{CARLA}} > 2 \sigma $. Table \ref{tablecarla_irac1} lists the results for all redshifts for both the $\alpha$ free and $\alpha$ fixed to $\alpha = -1$. The background subtraction and fitting procedure are identical to the analysis for IRAC2 detected sources (see Section 3). Similar to the results of the $4.5\ \mu$m Schechter fits, the IRAC1 luminosity functions are well described by $\alpha = -1$. Note that due to the shallower IRAC1 observations, the uncertainties on $\alpha$ for the free $\alpha$ fits are about two times larger than for the IRAC2 luminosity functions. Therefore, the fixed-$\alpha$-fits probably constrain $m^{*}_{3.6\mu m}$ better than the free-$\alpha$ fits. 

\begin{table}
\caption{Schechter fit results for both $\alpha$ free and $\alpha$ fixed fits to the 3.6$\mu$m luminosity function.}
\label{tablecarla_irac1}
\begin{center}
\begin{tabular}{l c |c c| c |c c}
\hline\hline
$\Sigma_{\rm{CARLA}}$ & $\langle{z}\rangle$ & m$^*_{3.6\mu m}$ & $\alpha$ & m$^*_{3.6 \mu m, \alpha = -1}$ & N  \\
\hline
      $> $2 $\sigma$ &1.45       & $20.30       ^{+ 0.30      }_{-      0.20    }$ & $-0.42       ^{+ 0.25      }_{- 0.20     } $&$19.62       ^{+ 0.25      }_{- 0.20       }$&28\\
      $> $2 $\sigma$ &1.77       & $20.05       ^{+ 0.30      }_{-      0.20    }$ & $-1.14       ^{+ 0.20      }_{- 0.20     } $&$20.22       ^{+ 0.25      }_{- 0.20       }$&30\\
      $> $2 $\sigma$ &2.05       & $19.98       ^{+ 0.30      }_{-      0.20    }$ & $-0.99       ^{+ 0.20      }_{- 0.15     } $&$19.97       ^{+ 0.25      }_{- 0.20       }$&30\\
      $> $2 $\sigma$ &2.26       & $19.64       ^{+ 0.25      }_{-      0.25    }$ & $-1.30       ^{+ 0.20      }_{- 0.20     } $&$20.14       ^{+ 0.25      }_{- 0.25       }$&29\\
      $> $2 $\sigma$ &2.51       & $20.03       ^{+ 0.30      }_{-      0.20    }$ & $-0.91       ^{+ 0.25      }_{- 0.20     } $&$19.92       ^{+ 0.25      }_{- 0.20       }$&30\\
      $> $2 $\sigma$ &2.92       & $19.70       ^{+ 0.30      }_{-      0.25    }$ & $-1.06       ^{+ 0.25      }_{- 0.20     } $&$19.79       ^{+ 0.30      }_{- 0.20       }$&30\\
 \end{tabular}
\end{center}
\end{table}

\clearpage

\end{document}